\documentclass[12pt,onecolumn,journal]{IEEEtran}
\usepackage{latexsym}
\usepackage{amsmath,amssymb,graphicx}% ,makeidx}
\usepackage{citesort}
 \usepackage{algorithm}
    \usepackage{algorithmic}

\long\def\comment#1{}

\newfont{\bbb}{msbm10 scaled 700}

\newfont{\bb}{msbm10 scaled 1100}
\newcommand{\CC}{\mbox{\bb C}}

\newcommand{\EE}{\mbox{\bb E}}

\newcommand{\wv}{{\bf w}}

\newcommand{\xv}{{\bf x}}
\newcommand{\yv}{{\bf y}}

\newcommand{\Hm}{{\bf H}}

\newcommand{\Xm}{{\bf X}}

\newcommand{\Cc}{{\cal C}}

\newcommand{\Nc}{{\cal N}}

\newcommand{\Sc}{{\cal S}}

\newcommand{\alphav}{\hbox{\boldmath$\alpha$}}

\renewcommand{\det}{{\hbox{det}}}
\newcommand{\trace}{{\hbox{tr}}}

\newcommand{\herm}{{\sf H}}

\input{epsf}
\usepackage{verbatim}

\newtheorem{thm}{Theorem}[section]
\newtheorem{lemma}{Lemma}[section]
\newtheorem{cor}{Corollary}[section]

\newtheorem{claim}{Claim}[section]

\renewcommand{\baselinestretch}{1.3}

\begin{document}
\title{On the Distortion SNR Exponent of Some Layered Transmission Schemes}
\author{\authorblockN{Kapil Bhattad\authorrefmark{1},
Krishna Narayanan \authorrefmark{1}, and Giuseppe Caire \authorrefmark{2}}
\\\authorblockA{\authorrefmark{1}Texas A \& M University, College Station,
TX 77843. kbhattad,krn@ece.tamu.edu}
\\\authorblockA{\authorrefmark{2}
%Dept. of Electrical Engineering,Viterbi School of Engineering,
University of Southern California, Los Angeles, CA 90089. caire@usc.edu}
}

\maketitle
\begin{abstract}
We consider the problem of joint source-channel coding for
transmitting $K$ samples of a complex Gaussian source over $T = bK$
uses of a block-fading multiple input multiple output (MIMO) channel
with $M$ transmit and $N$ receive antennas. We consider the case
when we are allowed to code over $L$ blocks. The channel gain is
assumed to be constant over a block and channel gains for different
blocks are assumed to be independent. The performance measure of
interest is the rate of decay of the expected mean squared error
with the signal-to-noise ratio (SNR), called the distortion
SNR exponent. We first show that using a broadcast strategy as in
\cite{erkip-IT}, but with a different power and rate allocation
policy, the optimal distortion SNR exponent can be achieved for
bandwidth efficiencies $0 \leq b < (|N-M|+1)/\min(M,N)$. This is
the first time the optimal exponent is characterized for
$1/\min(M,N) < b < (|N-M |+ 1)/ \min(M , N )$. Also, for $b > MNL^2$, we show
that the broadcast scheme achieves the optimal exponent of $MNL$.
Special cases of this result have been derived in \cite{erkip-IT}
for the $L=1$ case and in \cite{erkip-ISIT} for $M=N=1$. We then
propose a digital layered transmission scheme that uses both time
layering and superposition. This includes many known schemes in
\cite{erkip,erkip-IT} as special cases. The proposed scheme is at
least as good as the currently best known schemes for the entire
range of bandwidth efficiencies, whereas at least for some $M$,
$N$, and $b$, it is strictly better than the currently best known
schemes.
\end{abstract}

\section{Introduction}
\subsection{Problem Statement}

Consider the problem of transmitting $K$ samples of a complex
Gaussian source in $T = bK$ uses of an $M \times N$ MIMO channel
with block fading where $b$ is the ratio of the channel bandwidth to
the source bandwidth.  The channel is given by
\begin{equation} \label{mimo-channel}
\yv_t = \sqrt{\frac{\rho}{M}} {\bf H}_{\lceil\frac{Lt}{T}\rceil}
\xv_t + \wv_t, \;\;\; t = 1,\ldots, T
\end{equation}
where: $T$ is the duration (in channel uses) of the transmitted
block; ${\bf H}_l \in \CC^{N \times M}$, $l = 1,\ldots,L$, is the
channel matrix for $\frac{(l-1)T}{L} < t \leq \frac{lT}{L}$
containing random i.i.d. elements $h^{l}_{i,j} \sim \Cc\Nc(0,1)$
(Rayleigh independent fading). The channel matrix for different
blocks are independent; $\xv_t$ is the transmitted signal at time
$t$; the transmitted codeword, $\Xm = [\xv_1,\ldots,\xv_T]$, is
normalized such that $\trace(\EE[\Xm^\herm\Xm]) \leq MT$; $\wv_t
\sim \Cc\Nc(0,I^{M \times M})$ is additive white Gaussian noise;
$\rho$ denotes the Signal-to-Noise Ratio (SNR), defined as the ratio
of the average received signal energy per receiving antenna to the
noise per-component variance. We also define $m = \min\{M,N\}$ and
$n = \max\{M,N\}$.

When the channel state information is available at both the
transmitter and the receiver, Shannon's separation theorem applies
and separate source and channel coding is optimal. However, when the
channel state information is available only at the receiver, the
separation theorem fails and the optimal scheme requires joint
source and channel coding.

Consider a family of joint source-channel coding schemes
$\{\Sc\Cc_b(\rho)\}$ of spectral efficiency $1/b$, where
$\Sc\Cc_b(\rho)$ denotes the scheme that operates at SNR $\rho$.
Corresponding to the coding scheme
$\Sc\Cc_b(\rho)$, let $D(\rho)$ denote the distortion averaged over the source, the noise,
and the channel realization.
The {\em distortion SNR exponent} of the family is defined as the
limit
\begin{equation} \label{snr-exponent}
a(b) = - \lim_{\rho\rightarrow \infty} \frac{\log D(\rho)}{\log
\rho}.
\end{equation}
The distortion SNR exponent {\em of the channel}, denoted by
$a^\star(b)$, is the supremum of $a(b)$ over all possible coding
families. We are interested in characterizing $a^\star(b)$ for the
block fading MIMO channel.

\subsection{Prior Work}
The diversity multiplexing tradeoff \cite{zheng-tse} is closely
related to the problem considered here. In \cite{zheng-tse}, Zheng
and Tse consider the problem of transmitting digital information
over a MIMO fading channel. For a family of coding schemes
$C_r(\rho)$ whose rate grows as $r \log \rho$, where $r$ is referred
to as the multiplexing rate, the diversity order of the family is
defined as the limit
\begin{equation}
d(r) = - \lim_{\rho \rightarrow \infty} \frac{\log P_e(\rho)}{\log
\rho}
\end{equation} where $P_e(\rho)$ denotes the probability of decoding error
corresponding to the coding scheme $C(\rho)$. The diversity order of
the channel, $d^{*}(r)$, is the supremum of $d(r)$ taken over all
possible coding families. In \cite{zheng-tse}, for the Rayleigh
fading channel, the diversity order was determined to be
\begin{equation}
d^{*}(r) = (M-k)(N-k) - (M+N-1-2k)(r-k) \label{eqn:mimodmt}
\end{equation}
 where $k = \lfloor r \rfloor$ for $0 < r < m$ and $0$ for $r>m$.

The distortion SNR exponent problem has been considered previously by
many researchers in
\cite{caire05,caire-IT,narayanan06,erkip,erkip-IT,erkip-ISIT,erkip-ITW,goldsmith2,goldsmith-holliday,laneman03,laneman-ISIT,laneman05}.
Distortion SNR exponent was first defined by Laneman {\em et al.} in
\cite{laneman03}. In \cite{laneman03,laneman-ISIT,laneman05} the
authors compared the performance of two schemes for parallel fading
channels (a) a separation based scheme and (b) a multiple
description based scheme where the message sent on each channel
corresponded to a description. If the multiplexing rate of the
channel code is low the probability of outage is low. However, the
corresponding quantization error is large. When the multiplexing
rate is increased quantization error decreases but outage
probability increases. For these schemes, the optimal
multiplexing rate is chosen such that it maximizes the distortion SNR exponent.
Goldsmith and Holliday \cite{goldsmith2,goldsmith-holliday} consider
a separation based scheme for the MIMO channel and derive the
optimal operating point (multiplexing rate of the channel code) that
maximizes the distortion SNR exponent.

An upper bound on $a^\star(b)$ was derived by Caire and Narayanan
\cite{caire05,caire-IT,narayanan06} and by Gunduz and Erkip
\cite{erkip,erkip-IT,erkip-ISIT,erkip-ITW} by  assuming that the
transmitter is
informed of the channel realization $\Hm = \{{\bf H}_1,\ldots,{\bf
H}_L\}$. In this case, Shannon's separation theorem applies and the
optimal distortion is given by $D(\Hm) = 2^{-2R(\Hm)}$ where $R(\Hm)
= \frac{1}{L}\sum_{l} \log \det (I+\frac{\rho}{M} {\bf H}_l {\bf
H}_l^H)$. The distortion SNR exponent is then the exponent corresponding
to $E_{\Hm}[D(\Hm)]$. This has been computed in closed form for the
Rayleigh fading channel in \cite{caire05,caire-IT,narayanan06,erkip,erkip-IT,erkip-ISIT,erkip-ITW} and
is given by
\begin{equation}
a_{IT}(b) =  \sum_{i=1}^{m} \min (b,(2i-1+n-m)L). \label{eqn:ub}
\end{equation}
 Note that this is an
upper bound and is not known to be achievable.

The schemes by Laneman {\em et al.}
\cite{laneman03,laneman-ISIT,laneman05} and Goldsmith and Holliday
\cite{goldsmith2,goldsmith-holliday} are far away from the informed
transmitter upper bound. In \cite{caire05,caire-IT,narayanan06}, two
hybrid digital analog (HDA) scheme were proposed for $b < 1/m$ and
$b > 1/m$. For $b < 1/m$, in the HDA scheme, the transmitted signal
was chosen to be a superposition of an analog layer with a digital
layer. The analog layer is formed by a fraction $mb$ of the source
symbols. The remaining source symbols were quantized and transmitted
in the digital layer. The scheme was shown to achieve the upper
bound for $b < 1/m$. For $b > 1/m$, the HDA scheme involved
transmitting in two ``time'' layers (i.e., two layers multiplexed in
time). A digital layer of bandwidth $b - 1/m$ ($T-K/m$ channel uses)
was used to transmit the quantized source and the quantization error
was transmitted in an analog layer of bandwidth $1/m$. This scheme
improved on the exponent obtained by the separation based scheme.
However, the gap to the upper bound was still large.

In \cite{erkip,erkip-ITW,erkip-IT}, Gunduz and Erkip proposed a
hybrid layering scheme (HLS) that improved on the exponent
obtained by the HDA scheme for $b > 1/m$ by allowing for multiple
digital time layers instead of the single digital layer of the HDA
scheme. They also proposed a broadcast scheme (BS) that involved
transmitting a superposition of several digital layers. For the
$L=1$ case, the broadcast scheme was shown to achieve an exponent
of $MN$ for $b > MN$ which overlaps with the upper bound and is
hence optimal. In this case, for the region $1/m < b < MN$, a
characterization of the best achievable distortion SNR exponent is
not available. Currently the best known exponents are obtained by
the hybrid layering scheme  and broadcast strategy of Gunduz and
Erkip \cite{erkip-IT}. In \cite{erkip-ISIT}, Gunduz and Erkip
considered the broadcast scheme for parallel channels which
corresponds to $M = N = 1$ and $L>1$ in our model and they showed
that the broadcast scheme achieves an exponent of $MNL$ for $b
> L^2$ and is hence optimal. Note that throughout this paper we
refer to a superposition coding scheme as a broadcast scheme.

In other related work, Dunn and Laneman \cite{laneman-allerton} consider the
distortion to be of the form
\begin{equation}
D \approx C(b) \log (b \rho)^p \rho^{-a(b)}
\end{equation}
and compare several schemes using this approximation.

\subsection{Main Results}

The main results presented in this paper are summarized below.
\begin{enumerate}
\item We fully characterize the exponent achievable by any
broadcast (superposition) scheme. An achievable exponent and the
corresponding rate and power allocation are specified in
Theorem~\ref{thm:broadcast}. In Theorems~\ref{thm:broadcastUB1}
and \ref{thm:broadcastUB2}, we show that no broadcast scheme can
outperform the scheme in Theorem~\ref{thm:broadcast}.

\item We show that the broadcast scheme in \cite{erkip-IT} when
used with a different power and rate allocation than that
specified in \cite{erkip-IT} achieves the optimal exponent $mb$
for $b < \frac{n-m+1}{m}$. \item The broadcast scheme with the
proposed power and rate allocation policy achieves the optimal
exponent of $MNL$ for $b > MNL^2$. Special cases of this result
have been derived in \cite{erkip-IT} for the $L=1$ case and in
\cite{erkip-ISIT} for $M=N=1$. \item The proposed power and rate
allocation policy for the broadcast scheme becomes identical to
that specified in \cite{erkip-IT} for $MNL - (M+N-1)L < b < MNL -
(M+N-1)(L-1)$.  For other $b < MNL^2$ the distortion SNR exponent
obtained is larger than the broadcast scheme exponent of
\cite{erkip-IT}. \item We propose a time layering scheme in which
the last time layer is a broadcast layer, i.e, the last time layer
is a superposition of several layers. The distortion SNR exponent
obtained using this scheme is shown to be better than the exponent
obtained using the HLS scheme of \cite{erkip-IT}. We refer to this
scheme as LSBLEND as an abbreviation for Layered Scheme with a
Broadcast Layer at the end. \item We also propose a layering
strategy, termed the Box scheme, which generalizes BS and LSBLEND
proposed in this paper and the strategies considered earlier in
\cite{caire05,erkip-IT} by allowing for superposition and time
layers simultaneously. All previously known schemes are special
cases of the Box scheme and hence the Box scheme performs at least
as well as these schemes. However, the optimal distortion SNR
exponent for the Box scheme is difficult to obtain. We present a
suboptimal algorithm to compute an achievable distortion SNR
exponent. The scheme with the suboptimal algorithm is shown to
outperform all previously known schemes, including BS and LSBLEND
which are proposed in this paper, for some range of $b$, whereas,
for the considered examples, they are at least as good as
previously known schemes for all $b$.
\end{enumerate}
Some of these results have been reported in a conference version
of this paper \cite{bhattad-asilomar}.

\subsection{Organization of the paper}
The paper is organized as follows.
The proposed
schemes - Broadcast Scheme, LSBLEND, and Box Scheme, for the $L=1$
case, are discussed in section \ref{sec:bs}, \ref{sec:lsblend}, and
\ref{sec:box} respectively. The results for $L > 1$ case are
presented in section \ref{sec:generalL}. In section
\ref{sec:results}, we present some examples that demonstrate that
the proposed schemes achieve better distortion SNR exponent than
previously known schemes and finally we conclude in section
\ref{sec:conclusion}.

%\section{Preliminaries and Prior Results}
%In this section we briefly summarize the upper bound and
%the best known achievable region for the distortion exponent.
%\begin{lemma}Distortion exponent upper bound is given by
%  \begin{equation}
%    a^\star(b) \leq \sum_{i=1}^{m} \min(b,2i-1+n-m).
%  \end{equation}
%\end{lemma}

%\begin{lemma} The achievable distortion exponent with hybrid digital
%analog scheme is
%\begin{displaymath}
%  a^{BS}(b)= \left\{ \begin{array}{ll}
% b m & \textrm{for $b < \frac 1 m$}\\
% m n & \textrm{for $b \geq \frac 1 m$}
%  \end{array} \right.
%\end{displaymath}
%This is optimal for $b < 1/m$.
%\end{lemma}

%\begin{lemma}Achievable distortion exponent with broadcast strategy
%\begin{displaymath}
%  a^{BS}(b)= \left\{ \begin{array}{ll}
% b  & \textrm{for $b < mn$}\\
% m n & \textrm{for $b \geq mn$}
%  \end{array} \right.
%\end{displaymath}
%The broadcast strategy is optimal for all $b$ when $m = 1$
%and for $b > mn$ when $m > 1$.
%\end{lemma}

\section{Proposed Schemes}
In this section we present the proposed schemes
for the L=1 case. The results for the $L>1$ case are presented in section
\ref{sec:generalL}.

%In this section we present the broadcast scheme (sec. \ref{sec:bs}),
%the layered scheme with broadcast layer at the end (sec.
%\ref{sec:lsblend}), and the box scheme (sec. \ref{sec:box}). We
%restrict ourself to the $L=1$ case. The distortion exponent for the
%$L>1$ case can be computed in a similar manner as the $L=1$ case.
%The results for the $L>1$ case are presented in sec.
%\ref{sec:generalL}.

\subsection{Digital Layering using Superposition Only}
\label{sec:bs}

\begin{figure}[h]
\begin{center}
\includegraphics*[width=5in]{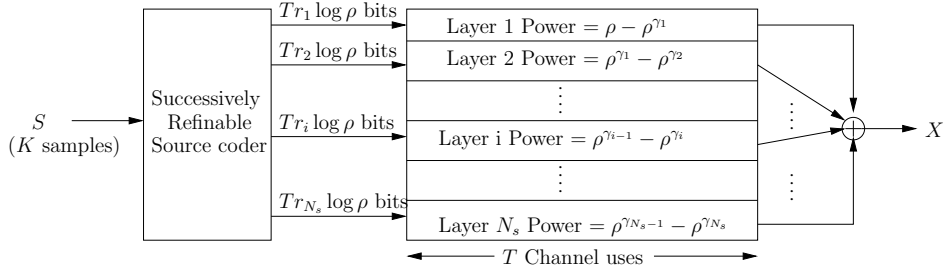}
\caption{Broadcast Schemes} \label{fig:bs}
\end{center}
\end{figure}

Consider the broadcast scheme shown in Fig. \ref{fig:bs}. The scheme
has $N_s$ superposition layers with the $i$th superposition layer
being assigned a power level of $\rho^{\gamma_{i-1}}
-\rho^{\gamma_{i}}$ where $\rho$ is the signal-to-noise ratio and
$\gamma_i \geq 0$  is a decreasing sequence with $\gamma_0 = 1$. The
source is compressed into $N_s$ layers such that it is successively
refinable. The rate in the $i$th refinement layer is $\frac{T r_i \log
\rho}{K} = b r_i \log \rho$.
Therefore, if a receiver estimates the source using the first $J$
layers the resulting distortion would be $2^{-\sum_{i=1}^{J}b r_i
\log \rho } = \rho ^{-b \sum_{i=1}^{J} r_i}$. The $i$th refinement
layer is transmitted in the $i$th superposition layer. Since $T r_i
\log \rho$ bits have to be transmitted in $T$ uses of the channel,
the resulting rate of transmission in the $i$th broadcast layer is
$r_i \log \rho$. For mathematical convenience in deriving the
expressions, we will assume that in the last layer (layer $N_s+1$)
the remaining power of $\rho ^ {\gamma_{N_s}}$ is used to transmit
Gaussian noise. Therefore, $\gamma_{N_s+1} = 0$ and $r_{N_s+1}=0$.
The channel codes used in the broadcast layers are assumed to
achieve the diversity multiplexing tradeoff \cite{zheng-tse}
corresponding to that layer. Here achieving the diversity
multiplexing tradeoff refers to achieving an error probability that
decays as $\rho^{-d(r)}$ with a coding rate that grows as $r \log
\rho$, where $d(r)$ is the optimal diversity multiplexing tradeoff
function corresponding to that layer.

At the receiver, the decoder attempts to decode as many layers as it
can using successive interference cancellation starting from the
first layer. That is, it decodes layer 1 by treating the signal
transmitted in layers $2$ to $N_s$ as noise. On successful decoding
it removes the contribution of layer 1 from the received signal and
repeats the process for layer 2 and so on. It then makes an estimate
of the source using all the layers it is able to decode.

To compute the distortion SNR exponent of the broadcast scheme,
we first characterize the diversity multiplexing tradeoff
of the broadcast scheme in the following lemma.

\begin{lemma}
\label{lemma:dstar}
If the multiplexing gain in the $i$th layer of the broadcast
scheme is $r_i = k (\gamma_{i-1} - \gamma_{i}) + \delta $
where $k \in \{0,1,\ldots,m-1\}$ and
$0 \leq \delta < \gamma_{i-1} - \gamma_{i}$, $\gamma_{i-1} > \gamma_{i} \geq 0$,
then, the achievable diversity in the $i$th layer of the
broadcast scheme, assuming that the message transmitted in the
previous layers is available at the receiver, is given by
\begin{equation}
d^\star(r_{i},\gamma_{i-1},\gamma_{i}) = (m-k) (n-k) \gamma_{i-1} -
(m+n-1-2k) \delta. \label{eqn:dstar}
\end{equation}
That is, if
\begin{equation}
X = \frac{1}{\sqrt \rho}\left(\sum_{i=1}^{N_s}
\sqrt{(\rho^{\gamma_{i-1}} - \rho^{\gamma_i})}X_i+\sqrt{\rho^{\gamma_{N_s}}}N_1 \right),
\end{equation}
where $X_i,N_1 \sim {\cal CN}(0,I^{M \times M})$, is transmitted
over a MIMO channel $Y = \sqrt{\frac{\rho}{M}}\Hm X + N$, then the
probability of the outage event
\begin{equation}
{\cal A}_i = \{H : I(X_i;Y|\Hm = H, X_1,\ldots X_{i-1}) < r_i \log
\rho\}
\end{equation} is
given by $P({\cal A}_i) \doteq
\rho^{-d^\star(r_{i},\gamma_{i-1},\gamma_{i})}$. (Here $A \doteq B$
is used to denote that $A$ and $B$ are equal in exponential order,
i.e., $\lim_{\rho \rightarrow \infty} \frac{\log A}{\log \rho}
=\lim_{\rho \rightarrow \infty} \frac{\log B}{\log \rho}$.)
Note that the term $\sqrt{\rho^{\gamma_{N_s}}}N_1$ in $X$ is the
Gaussian noise transmitted in layer $N_s+1$ and is
introduced for mathematical convenience.
It should not be confused with noise from the channel.
\end{lemma}
\begin{proof}
\begin{equation}
P({\cal A}_i) = P\left(\log \frac{\det (I + \frac{1}{M}
\rho^{\gamma_{i-1}} \Hm \Hm^H)} {\det (I +
\frac{1}{M}\rho^{\gamma_i} \Hm \Hm^H)} < r_i \log \rho\right).
\end{equation}

Let $\lambda_1,\ldots,\lambda_m$ denote the non-zero ordered eigenvalues
of $\Hm \Hm^H$ with $\lambda_1 \leq \lambda_2 \leq \ldots
\lambda_m$. As in \cite{zheng-tse}, let $\alpha_j = - \frac{\log
\lambda_j}{\log \rho}$. Therefore, $\alpha_1 \geq \alpha_2 \geq
\ldots \alpha_m$. Then
\begin{equation}
P({\cal A}_i) = P \left(\log \prod_{j=1}^{m} \frac{1 +
\frac{1}{M}\rho^{\gamma_{i-1}-\alpha_j}}{1 +
\frac{1}{M}\rho^{\gamma_{i}-\alpha_j}} < r_i \log \rho\right).
\end{equation}

At high SNR, we obtain $P({\cal A}_i) \doteq P( {\cal A'}) $ where
\begin{equation}
{\cal A'}= \left\{\alpha: \sum_{j=1}^{m} (\gamma_{i-1} - \alpha_j)^+
- \sum_{j=1}^{m} (\gamma_{i}-\alpha_j)^+ < r_i \right\}.
\end{equation}

Starting from Lemma 3 of \cite{zheng-tse} and following in the
footsteps of \cite{zheng-tse} we obtain
\begin{equation}
    P({\cal A'}) = \int_{\cal A'} p(\alpha)d\alpha \doteq \int_{\cal A'\cap
\alphav^+} \prod_{j=1}^{m}
    \rho^{(2j-1+n-m)\alpha_j} d \alpha
\end{equation}
for the Rayleigh fading channel. Therefore the outage probability is
given by $P({\cal A}_i) \doteq
\rho^{-d^\star(r_i,\gamma_{i-1},\gamma_{i})}$ where
\begin{equation}\label{eqn:dstar1}
d^\star(r_i,\gamma_{i-1},\gamma_{i}) = \inf_{{\cal A'} \cap
\alphav^+} \sum_{j=1}^{m} (2j-1+n-m) \alpha_j.
\end{equation}

For $r_i = k(\gamma_{i-1}-\gamma_i) + \delta$ where $k \in
[0,1,\ldots,m-1]$ and $0 \leq \delta < \gamma_{i-1}-\gamma_i$, the
infimum in (\ref{eqn:dstar1}) occurs when $\alpha = \alpha^{*}$ where
\begin{equation}\alpha^{*}_j=\left\{
       \begin{array}{ll}
           \gamma_{i-1}, & 1 \leq j < m-k;\\
           \gamma_{i-1} - \delta, & j = m-k;\\
            0, &  m-k < j \leq m.
       \end{array}\right.
\end{equation}
%Note that since $\delta < \gamma_{i-1}-\gamma_i$ and $0 \leq
%\gamma_i \leq \gamma_{i-1}$, we have $0 \leq \delta < \gamma_{i-1}$.
%Therefore, $\alpha_{m-k} > 0$ and hence,
Substituting $\alpha^{*}$ in (\ref{eqn:dstar1}) we obtain
\begin{eqnarray}
\nonumber d^\star(r_i,\gamma_{i-1},\gamma_{i})
 &=& \left(\sum_{j=1}^{m-k-1} (2j-1+n-m)\right) \gamma_{i-1}
 + \left(2(m-k)-1+n-m\right) (\gamma_{i-1} - \delta)
\\ \nonumber
& = & \left(\sum_{j=1}^{m-k} (2j-1+n-m)\right) \gamma_{i-1} - \left(2(m-k)-1+n-m\right)
\delta
\\ \nonumber
& = & (m-k)(2\frac{m-k+1}{2}-1+n-m) \gamma_{i-1} - (m+n-1-2k)\delta.
\end{eqnarray}
This then gives the desired result.
\end{proof}

Note that the probability of the outage event
${\cal A}_i$ discussed in lemma \ref{lemma:dstar}
is different from (a) the outage probability of layer $i$
and (b) the outage probability of layer $i$ given layers
$1$ to $i-1$ are decoded. Rather, it is the probability of
outage of layer $i$ with a genie aided decoder where the
genie provides the signal that is transmitted in layers $1$ to $i-1$.

\begin{figure}
\begin{center}
\includegraphics*[width = 5in]{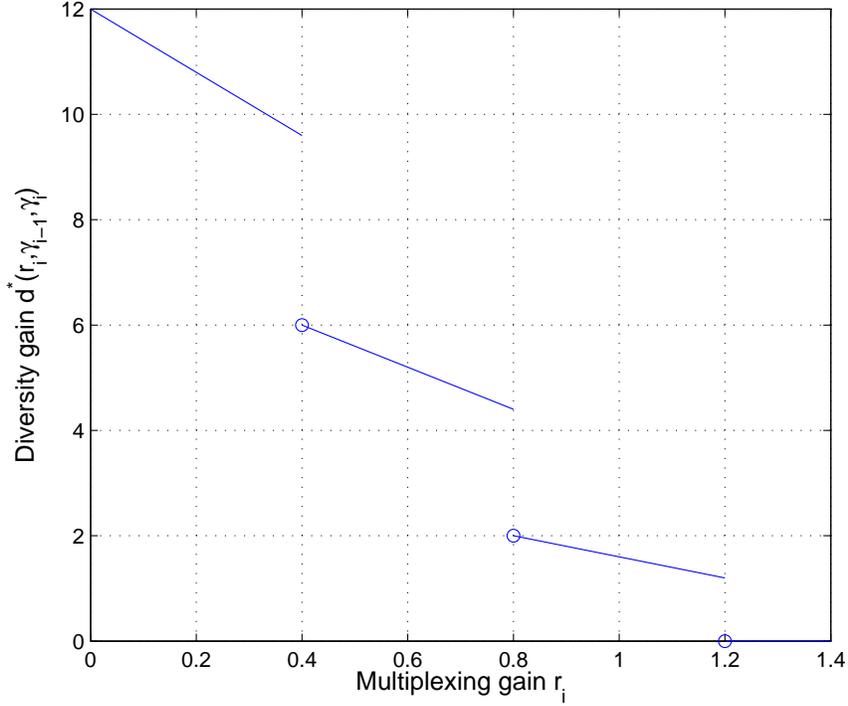}
\caption{Diversity Multiplexing tradeoff corresponding to a broadcast layer with $\gamma_{i-1}=1$
and $\gamma_{i}=0.6$ for a $3 \times 4$ MIMO system}
\label{fig:dstar}
\end{center}
\end{figure}

We will refer to the rate of decay of $P({\cal A}_i)$ with $\rho$,
i.e., $\lim_{\rho \rightarrow \infty} \frac{\log P({\cal A}_i)}{\log
\rho} = d^{*}(r_i,\gamma_{i-1},\gamma_i)$,  as the diversity of
layer $i$. In Fig. \ref{fig:dstar}, as an example, the diversity
multiplexing tradeoff corresponding to a superposition layer in the
broadcast scheme is plotted. Note that it is discontinuous.

Note that the diversity multiplexing tradeoff of Zheng and Tse \cite{zheng-tse}
specified in (\ref{eqn:mimodmt}) corresponds to the case when $\gamma_{i-1} = 1$ and $\gamma_{i} = 0$. From
Lemma \ref{lemma:dstar} and (\ref{eqn:mimodmt}) we can verify
that  $d^{*}(r_i, 1, 0)  = d^{*}(r_i)$. To keep the notation brief,
in such cases, we will use $d^{*}(r_i)$ in place of  $d^{*}(r_i, 1, 0)$.

The broadcast scheme considered in \cite{erkip-IT} used $r_i =
\gamma_{i-1} - \gamma_{i}$ and optimized the power allocation,
$\gamma_i$'s, in order to maximize the distortion SNR exponent. With
this rate and power allocation, the resulting scheme had a
distortion SNR exponent equal to $\min(b,MN)$.
We show that by using a different rate and power
allocation than that specified in \cite{erkip-IT}, we can obtain the
optimal exponent of $m b$ for any $b < \frac{n-m+1}{m}$. Notice that
in this region the broadcast scheme with the rate and power
allocation specified in \cite{erkip-IT} performs quite poorly. Our
main result in this section is the following theorem.

\begin{thm}
\label{thm:broadcast} The broadcast scheme achieves a distortion
SNR exponent of $(k+1)b$, $k \in \{0,1,\ldots,m-1\}$ for $
\frac{(M-k-1)(N-k-1)}{k+1} < b < \frac{(M-k)(N-k)}{k+1}$
with power and rate allocation
\begin{equation}
\gamma_i  = \left(\frac{b(k+1)
-(M-k-1)(N-k-1)}{(M-k)(N-k)-(M-k-1)(N-k-1)}\right)^{i}
\end{equation}
and
\begin{equation}
r_i =  (k+1) (\gamma_{i-1} - \gamma_{i}) - \epsilon
\end{equation}
for arbitrarily small $\epsilon > 0$.
\end{thm}
\begin{proof}
%We will use $D1 \doteq D2$ to denote that
%$D1$ and $D2$ are equal up to the exponential order, i.e.,
%$\lim_{ \rho \rightarrow \infty}
% \frac{\log D1}{\log \rho}  = \lim_{ \rho \rightarrow \infty} \frac{\log D2}{\log \rho}$.
The distortion is given by
\begin{equation}
D = \sum_{i=1}^{N_s} P(\mbox{Layer $1$ to $i-1$ decoded, layer
$i$ decoding failed}) D_i + P(\mbox{All layers decoded}) D_{N_s+1}
\label{eqn:D}
\end{equation}
where $D_i$ is the distortion when only the first $i-1$ layers
are used for reconstructing the source.
If the layers $1,\ldots,i-1$ can
be decoded, a source coding rate of $b \sum_{j=1}^{i-1}r_j$ can be obtained.
Therefore $D_i = \rho^{-b \sum_{j=1}^{i-1}r_j}$.

We have
\begin{eqnarray}
\nonumber \lefteqn{P(\mbox{Layer $1,\ldots ,i-1$ decoded, layer $i$
decoding failed})}\\ \nonumber
 &=& P(\mbox{Layer $1,\ldots ,i-1$ decoded, layer $i$ decoding failed} \mid X_1, \ldots, X_{i-1} \mbox{ available to decode layer $i$})\\ \nonumber
&\leq& P(\mbox{Layer $i$ decoding failed} \mid X_1,\ldots,X_{i-1} \mbox { available to decode layer $i$})\\
&\doteq&\rho^{-d^\star(r_{i},\gamma_{i-1},\gamma_{i})}.
\label{eqn:pouti}
\end{eqnarray}

If $d^\star(r_{i},\gamma_{i-1},\gamma_{i}) > 0$ for all $i$, then,
\begin{equation} P(\mbox{All layers decoded}) = 1 - \sum_i
P(\mbox{Layer $1,\ldots ,i-1$ decoded, layer $i$ decoding failed}
\doteq \rho^{0}. \label{eqn:poutns}
\end{equation}

From (\ref{eqn:D}), (\ref{eqn:pouti}), and (\ref{eqn:poutns}) we
have
\begin{equation}
D \doteq \sum_{i=1}^{N_s} \rho^{-(b\sum_{j=1}^{i-1}r_j +
d^{*}(r_i,\gamma_{i-1},\gamma_{i}))} + \rho^{-b
\sum_{i=1}^{N_s}r_i}. \label{eqn:overallD}
\end{equation}

Let
\begin{equation}
a(i) = b\sum_{j=1}^{i-1}r_j +
d^{*}(r_i,\gamma_{i-1},\gamma_{i}) \label{eqn:ithexponent}
\end{equation}
be the exponent corresponding to the case when the $i$th layer is in outage
and $a(N_s+1) = b \sum_{i=1}^{N_s} r_i$ the exponent when all layers are decoded.
From (\ref{eqn:overallD}), the distortion SNR exponent for the broadcast
scheme is
\begin{equation}
a_{BS}(b) = \max_{r,\gamma} \min_{i} a(i).
\end{equation} In the following proof, we
fix $r_i = (k+1) (\gamma_{i-1} - \gamma_{i})-\epsilon$
and optimize the
power allocation $\gamma_i$'s for $i = 1$ to $N_s$ in order to
maximize the exponent. Note that $\gamma_0 = 1$.
%Since the $r_i$'s
%are fixed in a particular manner, we obtain only an achievable exponent.

In section \ref{sec:appBS} of the Appendix, using the
Karush-Kuhn-Tucker (KKT) conditions, it is proved that for $b >
\frac{(M-k-1)(N-k-1)}{k+1}$ the optimal exponent is obtained when
all the exponents $a(i)$ are equal provided that the resulting solution
satisfies $\gamma_i > \gamma_{i+1}$ for all $i$ and $\gamma_{N_s} >
0$.

In order for $a(i)=a(i+1)$, from (\ref{eqn:ithexponent}) we have
\begin{equation} d^{*}(r_{i},\gamma_{i-1},\gamma_{i}) = b r_i +
d^{*}(r_{i+1},\gamma_{i},\gamma_{i+1}). \label{eqn:aiaiplus1}
\end{equation}
Since $r_i$ is chosen to be $(k+1) (\gamma_{i-1} -
\gamma_{i})-\epsilon$, from (\ref{eqn:dstar}) we have
\begin{eqnarray}
d^{*}(r_i,\gamma_{i-1},\gamma_i)= (M-k)(N-k)\gamma_{i-1}
 -(M+N-1 -2k)(\gamma_{i-1}-\gamma_i-\epsilon).
%\\&=& (n-m+1) \gamma_{i} + O(\epsilon).
\label{eqn:dstarthm1}
\end{eqnarray}
Substituting  from (\ref{eqn:dstarthm1}) in (\ref{eqn:aiaiplus1})
and using $r_i = (k+1)(\gamma_{i-1}-\gamma_i)$ we
have
\begin{eqnarray}
\nonumber \lefteqn{(M-k)(N-k)\gamma_{i-1} -
(M+N-1-2k)(\gamma_{i-1}-\gamma_{i})}
\\ \nonumber
&& = b(k+1)(\gamma_{i-1}-\gamma_{i}) + (M-k)(N-k)\gamma_{i} -
(M+N-1-2k)(\gamma_{i}-\gamma_{i+1}) + O(\epsilon).
\end{eqnarray}
On simplifying we obtain
\begin{equation} (\gamma_{i}-\gamma_{i+1})
= \alpha (\gamma_{i-1}-\gamma_{i}) + O(\epsilon)
\label{eqn:deltagamma}
\end{equation}
where \begin{equation} \alpha = \frac{b(k+1)
-(M-k-1)(N-k-1)}{M+N-1-2k}. \label{eqn:alpha}
\end{equation}
We can use (\ref{eqn:deltagamma}) recursively to obtain
\begin{equation}\label{eqn:gammadiff}
\gamma_i - \gamma_{i+1} = \alpha^i (\gamma_0 - \gamma_1) +
O(\epsilon)= \alpha^i (1 - \gamma_1)+ O(\epsilon).
\end{equation}
Therefore,
\begin{eqnarray}
1-\gamma_{i} &=& \sum_{j=1}^i (\gamma_{j-1} - \gamma_{j}) =
\sum_{j=1}^{i}\alpha^{j-1} (1-\gamma_{1}) + O(\epsilon)=\frac{1 -
\alpha ^{i}}{1-\alpha} (1-\gamma_{1}) + O(\epsilon).
\label{eqn:finalgammaeqn}
\end{eqnarray}

Furthermore, if $b\sum_{j=1}^{N_s}r_j = a(1)$, we have
\begin{eqnarray}
\nonumber b \sum_{j=1}^{N_s} (k+1)(\gamma_{j-1} -\gamma_{j})&=&
(M-k)(N-k) - (M+N-1-2k)(1-\gamma_{1})
  + O(\epsilon)\\ \nonumber
\Rightarrow b(k+1) \frac{1-\alpha^{N_s}}{1-\alpha} (1-\gamma_1)&=&
(M-k)(N-k) - (M+N-1-2k)(1-\gamma_{1})
  + O(\epsilon)\\ \nonumber
\Rightarrow
(1-\gamma_1)&=&\frac{(M-k)(N-k)(1-\alpha)}{b(k+1)(1-\alpha^{N_s})+(M+N-1-2k)(1-\alpha)}.
\end{eqnarray}
From (\ref{eqn:alpha}) we have
\begin{equation}
(1-\gamma_1) =
\frac{(M-k)(N-k)(1-\alpha)}{(M-k)(N-k)-b(k+1)\alpha^{N_s}}.
\label{eqn:finalgamma1}
\end{equation}
From (\ref{eqn:finalgammaeqn})
\begin{eqnarray}
(1-\gamma_i) &=&
\frac{(M-k)(N-k)(1-\alpha^i)}{(M-k)(N-k)-b(k+1)\alpha^{N_s}}
 \label{eqn:thm1oneminusgamma}
\\\gamma_i &=&  \frac{(M-k)(N-k) \alpha ^i -
b(k+1)\alpha^{N_s}}{(M-k)(N-k) - b(k+1)\alpha^{N_s}}.
\label{eqn:thm1gamma}
\end{eqnarray}

Consider the case when $0 \leq \alpha \leq 1$, i.e., when
$\frac{(M-k-1)(N-k-1)}{k+1} \leq b \leq \frac{(M-k)(N-k)}{k+1}$.
Since $1 \geq \alpha^i$ and since $(M-k)(N-k)/(b(k+1)) > 1 >
\alpha^{N_s}$, from (\ref{eqn:thm1oneminusgamma}) it follows that
$\gamma_i \leq 1$. From (\ref{eqn:thm1gamma}), since
$(M-k)(N-k)/(b(k+1)) > 1 > \alpha^{N_s-i}$, we have $\gamma_i \geq
0$ and we also observe that $\gamma_i$ is a decreasing sequence in
$i$. Therefore, this a valid power allocation.

The resulting exponent is
$b(k+1)\frac{(M-k)(N-k)(1-\alpha^{N_s})}{(M-k)(N-k)
-b(k+1)\alpha^{N_s}}$ and on taking the limit as $N_s \rightarrow
\infty$ we obtain $b(k+1)$.
\end{proof}

For the region $\frac{(M-k)(N-k)}{k+1} \leq b \leq \frac{(M-k)(N-k)}{k}$,
Theorem \ref{thm:broadcast} does not specify any achievable exponent.
But notice that the exponent corresponding to both $b = \frac{(M-k)(N-k)}{k+1}$
and $b = \frac{(M-k)(N-k)}{k}$ is $(M-k)(N-k)$. For this region,
we can ignore the additional bandwidth $b - \frac{(M-k)(N-k)}{k+1}$
and use a power allocation corresponding to $b = \frac{(M-k)(N-k)}{k}$ to
achieve an exponent of $(M-k)(N-k)$. The resulting achievable distortion
SNR exponent curve is continuous and is flat in the region
$\frac{(M-k)(N-k)}{k+1} \leq b \leq \frac{(M-k)(N-k)}{k}$ for $k = 1$ to $m-1$.
and for $b > MN$.

\begin{cor}
The optimal distortion SNR exponent for $b < (n-m+1)/m$ is $mb$.
\end{cor}
\begin{proof}
The result is obtained by comparing the upper bound in
%\cite{caire05}
(\ref{eqn:ub}) with the achievable exponent specified in Theorem
\ref{thm:broadcast} for the case when $k = (m-1)$.
\end{proof}

\begin{figure}
\begin{center}
\includegraphics*[width = 5in]{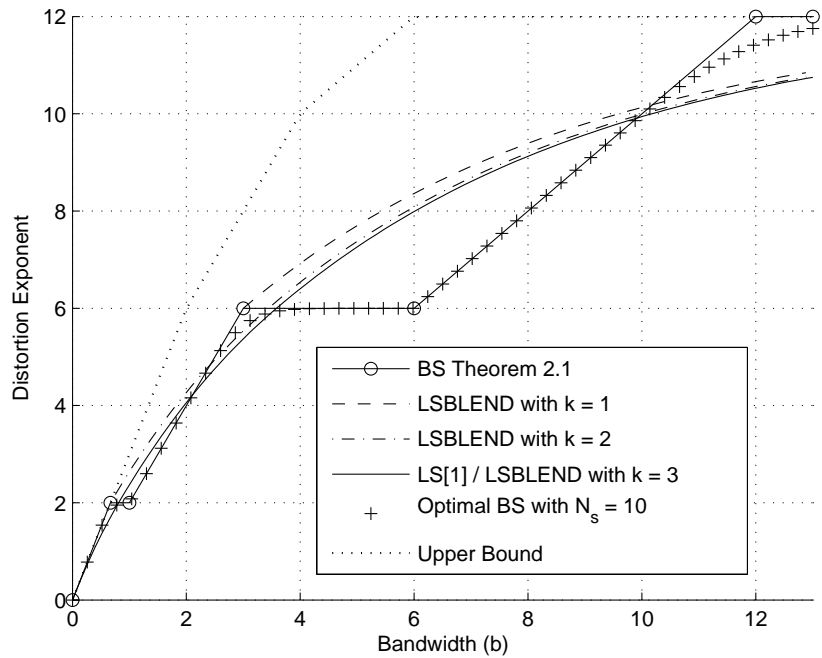}
\caption{Distortion SNR exponent for $M=3, N=4$}
\label{fig:results3x4}
\end{center}
\end{figure}

\begin{figure}
\begin{center}
\includegraphics*[width = 5in]{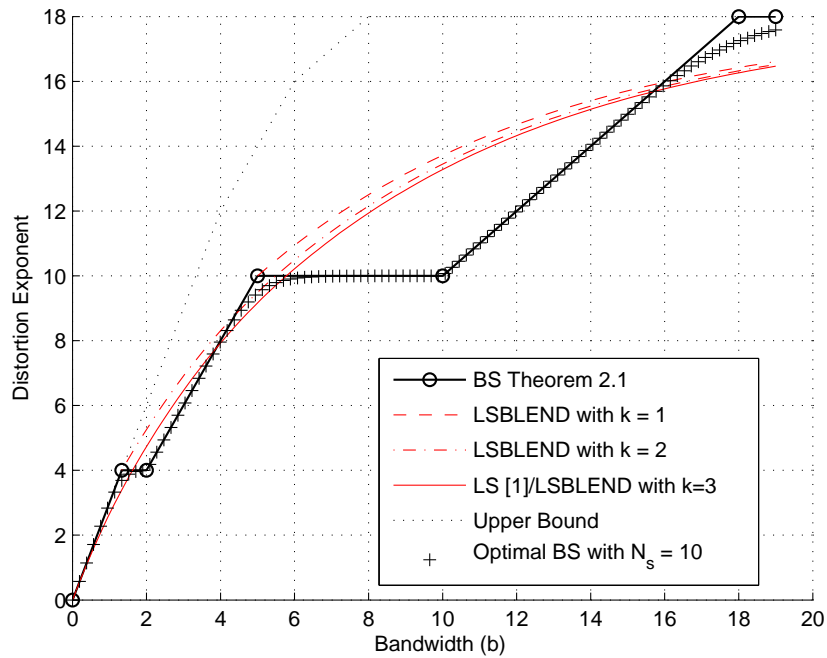}
\caption{Distortion SNR exponent for $M=3, N=6$}
\label{fig:results3x6}
\end{center}
\end{figure}

For $b < (n-m+1)/m$ and $b>mn$, BS achieves
the optimal exponent (it matches the informed transmitter upper bound)
and hence the power
and rate allocation specified in Theorem \ref{thm:broadcast}
is optimal.
%For $(n-m+1)/m < b < mn$, the rate and
%power allocation specified in Theorem \ref{thm:broadcast} may not be
%optimal. In the following two theorems we prove that the broadcast
%scheme can never achieve an exponent better than that specified in
%Theorem \ref{thm:broadcast} and hence the power allocation specified
%in Theorem \ref{thm:broadcast} is indeed optimal.
For the region between these two values, the next two results
prove that the exponent achieved in Theorem \ref{thm:broadcast} is
the optimal exponent achievable by any superposition (broadcast)
scheme. This is shown by finding an upper bound to the exponent of
any superposition scheme, for any power allocation and number of
layers, that  matches the achievable exponent of Theorem
\ref{thm:broadcast}. This also calls for schemes that are not
based on superposition alone in order to improve on the achievable
exponent in this region (discussed in the next sections).

\begin{thm}
For $b \leq \frac{(M-k)(N-k)}{k}$, the distortion SNR exponent of
the broadcast scheme $a_{BS}(b)  \leq (M-k)(N-k)$.
\label{thm:broadcastUB1}
\end{thm}
\begin{proof}
Recall that the exponent of the broadcast scheme is given by $a_{BS}(b) =
\min_i a(i)$ where $a(i)$ is as specified in
(\ref{eqn:ithexponent}).

Let us fix $b$ and $k$ such that $(M-k)(N-k)/k \geq b$.
Let us assume that there exists a power and rate allocation such that the exponent $a_{BS}(b)
> (M-k)(N-k)$, then since $a_{BS}(b) = \min_i a(i)$, then for all $i$ from $1$ to $N_s+1$
we must have $a(i) > (M-k)(N-k)$. As before, without loss of
generality, let the rate used in the $i$th layer be $r_i = k_i
(\gamma_{i-1} - \gamma_i) + \delta_i$, for some integer $k_i$ and
$0 \leq \delta_i < \gamma_{i-1} - \gamma_i$. We will now show that
if $a_{BS}(b) > (M-k) (N-k)$ were to be true, then $k_i < k$ for
all $i$.

%\begin{eqnarray*}
%&&a(1) > (M-k)(N-k)\\
%&&\Rightarrow (M-k_1)(N-k_1) - (M+N-1-2k_1)\delta_1
%>(M-k)(N-k)\\
%&&\Rightarrow k_1 \leq k-1.
%\end{eqnarray*}

The gist of the proof is as follows. If, to the contrary, $k_i
\geq k$ for some $i$, then there must be a smallest value of $i$
(say $i^*$) for which this is true. That is, there must be an $i^*
\geq 1$, for which $k_{i^*} \geq k$ and $k_i \leq k-1$ for all $i
= 1$ to $i^*-1$. We will now show that $a(i^*)$ cannot be larger
than $(M-k)(N-k)$.

We have
\begin{equation}
a(i^*) = b \sum_{i=1}^{i^*-1}r_i +
d^{*}(r_{i^*},\gamma_{i^*-1},\gamma_{i^*})
\end{equation}
Since, $r_i = k_i (\gamma_{i-1}-\gamma_i) + \delta_i$, clearly
$r_i \leq (k_i+1) (\gamma_{i-1}-\gamma_i)$. Therefore,
\begin{eqnarray*}
a(i^*) & \leq & b \sum_{i=1}^{i^*-1}(k_i+1)(\gamma_{i-1}-\gamma_i) + d^{*}(r_{i^*},\gamma_{i^*-1},\gamma_{i^*}) \\
& \leq & b \sum_{i=1}^{i^*-1}(k)(\gamma_{i-1}-\gamma_i) +  d^{*}(r_{i^*},\gamma_{i^*-1},\gamma_{i^*}) \ \ \ (\because k_i \leq k-1, \ {\mbox{for}} \ \ i < i^*)\\
& = & bk(1-\gamma_{i^*-1}) +  d^{*}(r_{i^*},\gamma_{i^*-1},\gamma_{i^*}) \\
& \leq & (M-k)(N-k)(1-\gamma_{i^*-1}) +
d^{*}(r_{i^*},\gamma_{i^*-1},\gamma_{i^*}) \ \ \  (\because b \leq (M-k)(N-k)/k)\\
& \leq & (M-k)(N-k)(1-\gamma_{i^*-1}) + (M-k_{i^*})(N-k_{i^*})\gamma_{i^*-1}\ \ \  (\because \delta_{i^*} \geq 0)\\
& \leq & (M-k)(N-k)(1-\gamma_{i^*-1}) + (M-k)(N-k)\gamma_{i^*-1}  \ \ \ (\because k_{i^*} \geq k)\\
&=&(M-k)(N-k).
\end{eqnarray*}

For $a_{BS} > (M-k) (N-k)$, we require $a(i) > (M-k) (N-k),
\forall i$ and, hence, we must have that $k_i \leq k-1$, for all
$i = 1, \ldots, N_s$. This implies that $r_i \leq
k(\gamma_{i-1}-\gamma_i)$. But, in this case,
\begin{equation*}
a(N_s+1) = b \sum_{i=1}^{N_s}r_i \leq b k (1 - \gamma_{N_s}) \leq
(M-k)(N-k).
\end{equation*}
Therefore, our assumption that $a_{BS}(b)$ can be greater than
$(M-k)(N-k)$ for $b < (M-k)(N-k)/k$ is not valid. Hence proved.
\end{proof}
As pointed out in the discussion after Theorem
\ref{thm:broadcast}, the achievable exponent for $(M-k)(N-k)/(k+1)
\leq b \leq (M-k)(N-k)/k$ is $(M-k)(N-k)$. This combined with the
upper bound specified in Theorem \ref{thm:broadcastUB1} proves
that this is the best achievable exponent using any broadcast
scheme for this range of $b$.

\begin{thm}
For $b > (M-k-1)(N-k-1)/(k+1)$ the distortion SNR exponent of the
broadcast scheme $a_{BS}(b)  \leq b(k+1)$. \label{thm:broadcastUB2}
\end{thm}
\begin{proof}
Recall that the exponent of the broadcast scheme is given by
$a_{BS}(b) = \min_i a(i)$ where $a(i)$ is as specified in
(\ref{eqn:ithexponent}). The idea of the proof is similar to that
in the proof of the previous theorem. Again we fix $b$ and $k$
such that $b > (M-k-1)(N-k-1)/(k+1)$. Let us assume that there
exists a power and rate allocation policy such that $a_{BS}(b) >
b(k+1)$. Let the rate allocation be $r_i = k_i
(\gamma_{i-1}-\gamma_i) + \delta_i$ as before.

The proof is similar to that of the previous theorem and is to
first show that $k_i \leq k$ for all $i$.
% we have
%\begin{eqnarray*}
%&&a(1) > b(k+1) > (M-(k+1))(N-(k+1)) \\
%&&\Rightarrow k_1 \leq k.
%\end{eqnarray*}
As before, let $i^* \geq 1$ be such that  Let $k_i \leq k$ for $i
= 1$ to $i^*-1$ and $k_{i^*} \geq k+1$. We have
\begin{eqnarray*}
a(i^*) &=& b \sum_{i=1}^{i^*-1}r_i + d^*(r_{i^*},\gamma_{i^*-1},\gamma_{i^*}) \\
&\leq& b (k+1) (1-\gamma_{i^*-1}) +
(M-k_{i^*})(N-k_{i^*})\gamma_{i^*-1} \ \ \
\\ && \qquad \qquad (\because r_i \leq (k_i+1)(\gamma_{i-1}-\gamma_i) \leq
(k+1)(\gamma_{i-1}-\gamma_i) \mbox{ for } i < i^*) \\
&\leq& b (k+1) (1-\gamma_{i^*-1}) + (M-k-1)(N-k-1)\gamma_{i^*-1} \
\ \
(\because k_{i^*} \geq k+1) \\
&=& b (k+1) - \gamma_{i^*-1}(b(k+1)-(M-k-1)(N-k-1)) \\
&\leq& b (k+1) \ \ \ (\because b > (M-k-1)(N-k-1)/(k+1)).
\end{eqnarray*}
This contradicts the assumption that $a_{BS}(b) > b(k+1)$.
Therefore, the only other possibility is that $k_i \leq k$ for all
$i$. In this case too, $a(N_s+1) = b \sum_{i=1}^{N_s} r_i \leq
b(k+1)$ which implies that the assumption $a_{BS}(b) > b(k+1)$ is
incorrect. Hence proved.
\end{proof}

Note that for $(M-k-1)(N-k-1)/(k+1) \leq b \leq (M-k)(N-k)/(k+1)$
the achievable exponent in Theorem~\ref{thm:broadcast} is also
$b(k+1)$. Hence, Theorem~\ref{thm:broadcast} along with
Theorems~\ref{thm:broadcastUB1} and \ref{thm:broadcastUB2} fully
characterize the exponent achievable with any broadcast scheme.

\underline{Finite Number of Layers:} In practice it is not
possible to have infinitely many layers and it is important to
study the performance of the broadcast scheme with a finite number
of layers. The problem of finding the optimal distortion SNR
exponent for a finite number of layers can be posed as the
following optimization problem.
\begin{eqnarray}
  &\max& a \\ \nonumber
  &\mbox{subject:}& \mbox{for all } i \in \{1,2,\ldots, N_s\} \\ \nonumber
&& \gamma_i \geq 0, \delta_i \geq 0, r_i \geq 0, k_i \in
\{0,1,\ldots, m-1\}, \\ \nonumber &&  \gamma_{i-1} > \gamma_{i} ,
\gamma_0 = 1, \\ \nonumber &&  \delta_i < \gamma_{i-1}-\gamma_i, \\
\nonumber &&  r_i = k_i(\gamma_{i-1} - \gamma_{i})+\delta_i, \\
\nonumber &&  a \leq b \sum_{j=1}^{i-1} r_j +
(m-k_i)(n-k_i)\gamma_{i-1}   -
(m+n-1-2k_i)\delta_i, \\ \nonumber &&  a \leq b \sum_{j=1}^{N_s}
r_j.
\end{eqnarray}
For a fixed set of $k_i$'s this reduces to a linear program.
For small $N_s$, the optimum exponent can be found by using
the linear program for all $m^{N_s}$ choices of $k_i$'s .

In Fig. \ref{fig:results3x4} and Fig. \ref{fig:results3x6}, the
distortion SNR exponent corresponding  to the broadcast scheme
proposed in Theorem \ref{thm:broadcast} is shown for a $3 \times 4$
and a $3 \times 6$ MIMO system. The optimal distortion SNR exponent
corresponding to the broadcast scheme with 10 layers is also shown.
We see that the exponent with finite layers is very close to the
best achievable distortion exponent of the broadcast scheme for all
$b$ and the curves overlap for a large range of $b$. Also as proved
in Theorem \ref{thm:broadcastUB1} and \ref{thm:broadcastUB2}, the
distortion exponent with finite layers does not improve on the
achievable exponent specified in Theorem \ref{thm:broadcast}.

\subsection{Layering in Time with one Broadcast layer at the end}
\label{sec:lsblend}

\begin{figure}[h]
\begin{center}
\includegraphics*[width=5in]{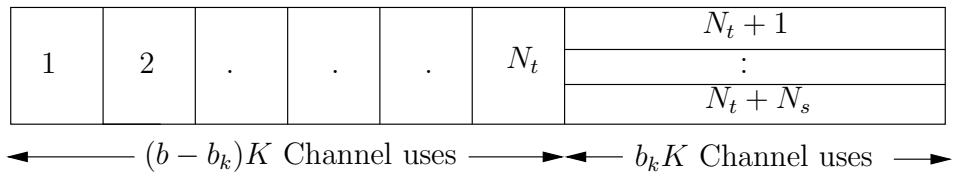}
\caption{Layered Schemes with Broadcast Layer at the end (LSBLEND)}
\label{fig:lsblend}
\end{center}
\end{figure}

Consider the scheme shown in Fig. \ref{fig:lsblend}. For $b
> b_k = (m-k)(n-k)/(k+1)$, $k \in \{1,\ldots,m\}$, a bandwidth of $b
- b_k$ is allocated for time layering and the remaining bandwidth of
$b_k$ is allocated to a broadcast scheme where the rate and power
allocation for the broadcast layers are chosen as specified in
Theorem \ref{thm:broadcast}. The parameter $k$ determines the
bandwidth splitting between the broadcast layer and the time layers.
The decoding proceeds by first decoding the time layers and then
decoding the broadcast layers after all the time layers are decoded.
This is similar to the HDA scheme of \cite{caire05} and the HLS of
\cite{erkip-IT} where the source is quantized and transmitted using
time layering in a bandwidth of $b - 1/m$  and the quantization
error is transmitted in an analog layer of bandwidth $1/m$. For the
proposed scheme, the distortion SNR exponent obtained for a particular
bandwidth splitting parameter $k$ is given in the following theorem.
The largest achievable distortion SNR exponent is then obtained by
taking a supremum over all $k$.

Note that when $k = m$, the bandwidth allocated to the broadcast
layer is 0, i.e., we have only time layering. This scheme, termed
Layered Scheme (LS), was proposed and analyzed in \cite{erkip-IT}.
The proof of the following theorem is similar to the derivation of
the exponent for LS in \cite{erkip-IT}.

\begin{thm}
Let $c_j = (m+n-1-2j) \log \frac{j+1}{j}$ for $j = 0, \ldots,m$. Let
$p$ be such that $\sum_{j=p+1}^{k-1} c_j \leq  (b-b_k) <
\sum_{j=p}^{k-1}c_j$. Then, the best distortion SNR exponent $a(b)$
achievable using LSBLEND is given by
\begin{equation}
a(b) =  mn-p-p^2- (m+n-1-2p)(p+1)e^{-\frac{b-b_k
-\sum_{j=p+1}^{k-1}c_j}{m+n-1-2p}}.
\end{equation}
\label{thm:lsblend}
\end{thm}
\begin{proof}
Let $N_t$ and $N_s$ denote the number of time and superposition
layers respectively. Let $a(i)$ for $i = 1,\ldots,N_t$ denote
the distortion SNR exponent corresponding to the case when the time layers
$1$ to $i-1$ are decoded and decoding of the $i$th time layer fails,
$a(N_t+i)$ for $i = 1,\ldots,N_s$ denote the distortion
SNR exponent corresponding to the case when all $N_t$ time layers and
the first $i-1$ broadcast layers are decoded while decoding of the
$i$th broadcast layer fails, and let $a(N_t+N_s+1)$ denote
the exponent corresponding to the case when all layers are decoded.

For the $i$th time layer, the probability of decoding failure is given by
$P_{e}(i) \doteq \rho^{-d^{*}(r_i)}$ where $r_i$ is the multiplexing rate
of the $i$th time
layer and $d^{*}(r_i)$ is the Zheng and Tse diversity multiplexing tradeoff function specified in (\ref{eqn:mimodmt}).
Note that power allocation to the time layer is $\rho^{1}-\rho^{0}$ and
$d^{*}(r_i) = d^{*}(r_i,1,0)$.
% = (m-k_i)(n-k_i) - (m+n-1-2k_i)\delta_i$
%where $k_i = \lfloor r_i \rfloor$ and $\delta_i = r_i - k_i$.
The bandwidth allocated to a time layer is
$(b-b_k)/N_t$. The distortion SNR exponent of the $i$th time layer is then given by
\begin{equation}
a(i) = \frac{b-b_k}{N_t} \sum_{j=0}^{i-1} r_j + d^{*}(r_i)
\end{equation}
where $r_0 = 0$.

For the broadcast layer we use the rate and power allocation as
specified in Theorem \ref{thm:broadcast}. With that rate and power
allocation it follows that the exponents
$a(N_t+1), a(N_t+2), \ldots a(N_t+N_s+1)$ are all equal and are given by
\begin{eqnarray}
\nonumber
a(N_t+i) &=& \frac{b-b_k}{N_t} \sum_{j=0}^{N_t} r_j +(k+1)b_k \\
 &=& \frac{b-b_k}{N_t} \sum_{j=0}^{N_t} r_j  + (m-k)(n-k)
\end{eqnarray}
for $i = 1$ to $N_s+1$ in the limit $N_s \rightarrow \infty$. Note
that we do not loose optimality here by fixing the rate and power
allocation of the broadcast layer since $(m-k)(n-k)$ is the maximum
possible contribution that the broadcast layer of bandwidth $b_k$
can make to the exponent (see Theorem \ref{thm:broadcastUB1} and
Theorem \ref{thm:broadcastUB2}).

In the following proof, we optimize $r_i$'s to maximize the
distortion SNR exponent.

In section \ref{sec:appLSBLEND} of the Appendix we show that the
exponent is maximized by choosing $r_i$'s such that $a(1) = a(2) =
\ldots  =a(N_t) =  a(N_t+1)$ provided that the resulting $r_i$'s lie
between $0$ and $m$. By setting $a(N_t) = a(N_t+1)$ we obtain
\begin{equation}
  d^{*}(r_{N_t}) = \frac{b-b_k}{N_t}  r_{N_t} + (m-k)(n-k).
\label{eqn:exponent_nt}
\end{equation}
We will consider the limiting case when $N_t \rightarrow \infty$.
From (\ref{eqn:exponent_nt}) we have, in the limiting case,
\begin{equation}
d^{*}(r_{N_t}) \rightarrow (m-k)(n-k).
\end{equation}
Therefore,
\begin{equation}
(m-k_{N_t})(n-k_{N_t}) - (m+n-1-2k_{N_t})\delta_{N_t} \rightarrow
(m-k)(n-k).
\end{equation}
This happens when $k_{N_t} = k-1$ and $\delta_{N_t} \rightarrow 1$.

By setting $a(i-1) = a(i)$ we have
\begin{equation}
  d^{*}(r_{i-1}) = \frac{b-b_k}{N_t}  r_{i-1} + d^{*}(r_i).
\label{eqn:exponent_eqn}
\end{equation}
$d^{*}(r)$ is a decreasing function and from
(\ref{eqn:exponent_eqn}) we have $d^{*}(r_{i-1}) \geq d^{*}(r_i)$.
Therefore $r_{N_t} \geq r_{N_t-1} \geq ... \geq r_1$. Let $r_{i}$
lie between $t$ and $t+1$. We want to check if $r_{i-1}$ also lies
between $t$ and $t+1$. To do so we assume that $k_{i-1}=t$ and solve
for $\delta_{i-1}$. If the resulting $\delta_{i-1}$ lies between $0$
and $1$, then the assumption $k_i = t$ is correct. From
(\ref{eqn:exponent_eqn}) we have
\begin{eqnarray}
\nonumber (m - t)(n-t) - (m+n-1-2t)\delta_{i-1}
 &=&  \frac{b-b_k}{N_t}(t+\delta_{i-1}) + (m - t)(n-t) - (m+n-1-2t)\delta_{i}.
\\\nonumber
\Rightarrow \delta_{i-1} (m+n-1-2t+\frac{b-b_k}{N_t} ) &=&
\delta_{i} (m+n-1-2t)-\frac{b-b_k}{N_t}t.
\\
\label{eqn:delta_recursion} \Rightarrow \delta_{i-1} &=& \alpha
\delta_{i} - (1-\alpha)t
\end{eqnarray}
where
\begin{equation}
\alpha =\frac{ m+n-1-2t }{m+n-1-2t + (b-b_k)/N_t} < 1.
\end{equation}

On using recursion (\ref{eqn:delta_recursion}) $N_l$ times we have
\begin{equation}
\delta_{i-N_l} = \alpha^{N_l} \delta_{i} - \frac{1 - \alpha^{N_l}}{1
-\alpha} (1-\alpha)t = \alpha^{N_l}(t+\delta_{i}) - t.
\label{eqn:afterdeltarec}
\end{equation}

The maximum number of times the recursion can be used such that the resulting
$\delta$ is positive is given by
\begin{eqnarray} \nonumber
\alpha^{N_l}(t+\delta_{i}) &\geq& t \\ \nonumber \Rightarrow N_l \log
\alpha &\geq& \log \frac{t}{t+\delta_{i}}\\ \nonumber \Rightarrow
\frac{N_l}{N_t} &\leq& \frac{1}{N_t \log \alpha }\log
\frac{t}{t+\delta_{i}} \qquad \because \alpha < 1 ,\ \log \alpha < 0
\\ \nonumber
\Rightarrow \frac{N_l}{N_t} &\leq& -\frac{1}{\log\left(\left(1 +
\frac{b-b_k}{(m+n-1-2t)N_t}\right)^{N_t} \right)}\log
\frac{t}{t+\delta_{i}}\\ \nonumber \Rightarrow \frac{N}{N_t} &\leq&
\frac{m+n-1-2t}{b-b_k}\log \frac{t+\delta_{i}}{t} \qquad (N_t
\rightarrow \infty).
\end{eqnarray}

For the proposed scheme, we start from $r_{N_t} = k$ ($k_{N_t}= k-1$
and $\delta_{N_t}=1$)
 and solve for $r_{i-1}$ from $r_{i}$.
Let $c_j = (m+n-1-2j) \log \frac{j+1}{j}$. If $p$ is such that $\sum_{j=p+1}^{k-1}
c_j \leq  (b-b_k) < \sum_{j=p}^{k-1}c_j$ then as $i$ decreases from
$N_t$, after a fraction $\sum_{j=p+1}^{k-1} c_j /(b-b_k)$ of the
time layers, $r_i$ decreases from $k$ to $p+1$. For the remaining
fraction $(1-\sum_{j=p+1}^{k-1} c_j / (b-b_k))$ of layers, as $i$
decreases, $r_i$ decreases but remains above $p$, i.e., $k_i$
remains constant at $p$ while $\delta_i$ decreases. From
(\ref{eqn:afterdeltarec}) we can calculate $r_1$ as
\begin{eqnarray}\nonumber
  r_1 &=& p + \lim_{N_t \rightarrow \infty}\alpha^{N_t(1 -\frac{1}{b-b_k}\sum_{j=p+1}^{k-1}c_j)}(p+1) - p\\
  &=& (p+1)e^{-(\frac{b-b_k}{m+n-1-2p})(1 -\frac{1}{b-b_k}\sum_{j=p+1}^{k-1}c_j)}.
\end{eqnarray}
The final exponent is given by
\begin{eqnarray} \nonumber
a(1) = d^{*}(r_1)
&=& (m-p)(n-p) - (m+n-1-2p)(r_1-p)\\ \nonumber
%&=& mn -p(m+n-p - (m+n-1-2p)) - (m+n-1-2p)r_1  \\\nonumber
&=& mn - p -p^2 - (m+n-1-2p)r_1
%&=& (m-p)(n-p)- (m+n-1-2p)((p+1)e^{-(\frac{b-b_0}{m+n-1-2p})(1 -\frac{1}{b-b_0}\sum_{j=p+1}^{M-2}c_j)}-p)\\ \nonumber
 %&=& mn-p-p^2- (m+n-1-2p)(p+1)e^{-(\frac{b-b_0}{m+n-1-2p})(1 -\frac{1}{b-b_0}\sum_{j=p+1}^{M-2}c_j)}\\ \nonumber
 %&=& mn-p-p^2- (m+n-1-2p)(p+1)e^{-\frac{b-b_0 -\sum_{j=p+1}^{M-2}c_j}{m+n-1-2p}}
\end{eqnarray}
which is the desired result.
\end{proof}
%For the special case when $m = 2$ we have $p = 0$ and the expression reduces to
%\begin{equation}
%exponent = 2n-(n+1)e^{-(\frac{b-b_0}{n+1})}
%\end{equation}

%Let $a_{LS}(b)$ as the exponent corresponding to LS (LSBLEND with
%$k=m$). Now consider the LSBLEND scheme with bandwidth splitting
%parameter $k$. If we replace the broadcast layer of bandwidth $b_k$
%whose contribution to the exponent is $b_k(k+1)$ by a layered scheme
%of bandwidth $a_{LS}^{-1}(b_k(k+1))$. Total bandwidth in the new
%scheme is $b-b_k+a_{LS}^{-1}(b_k(k+1))$. The contribution of the LS
%layer of bandwidth $a_{LS}^{-1}(b_k(k+1))$ to the distortion
%exponent is identical to that of the broadcast layer and therefore
%the rate allocation for the time layers in the bandwidth $b-b_k$
%remains the same. The new scheme is a layered scheme with bandwidth
%$b-b_k+a_{LS}^{-1}(b_k(k+1))$ and therefore we have
%\begin{equation}
%a_{LSBLEND,k}(b) = a_{LS}(b-b_k+a_{LS}^{-1}(b_k(k+1)))\ \ \mbox{for
%} b>b_k
%\end{equation}
%For the HLS scheme the exponent is given by
%$a_{LS}(b-1/m+a_{LS}^{-1}(1))$. When $m = n$ and $k = m-1$, the
%exponent form the LSBLEND scheme becomes identical to that of HLS.

Note that when $m = n$ and $k = m-1$, the contribution to the
distortion SNR exponent by the broadcast layer is $b_k(k+1) = 1$
and it uses a bandwidth of $b_k = 1/m$. In the HLS scheme, the analog layer
uses a bandwidth of $b_0 = 1/m$ and it also has a contribution of $mb_0  = 1$
towards the exponent. Therefore, in this case, the distortion SNR exponent
obtained with LSBLEND with $k = m-1$ is identical to that with HLS.
Therefore, the distortion SNR exponent obtained using LSBLEND
becomes identical to that obtained using HLS when (a) $m = n$ and
(b) the supremum occurs at $k = m-1$. It can be shown that
LSBLEND is strictly better otherwise for $b > 1/m$.

\subsection{Digital Layering in Time and Using Superposition}
\label{sec:box}

\begin{figure}[h]
\begin{center}
\includegraphics*[width=5in]{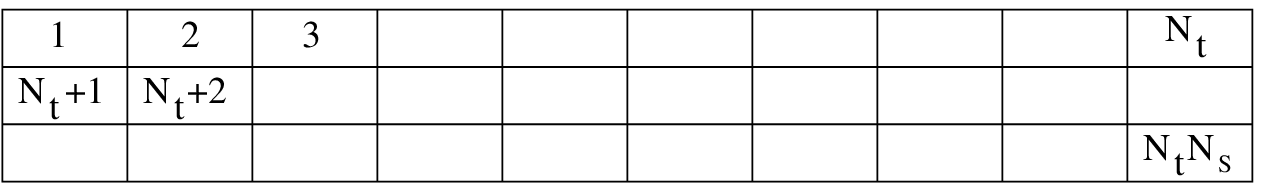}
\caption{Box Scheme} \label{fig:box}
\end{center}
\end{figure}

The source is encoded in such a way that it is successively
refinable. The transmitted signal composes of $N_t$ time layers
where each time layer is a superposition of $N_s$ layers. To the
$(i,j)$th layer, i.e., the $j$th time layer and the $i$th
superposition layer within it, we allocate a power level of
$\rho^{\gamma_{i-1,j}} - \rho^{\gamma_{i,j}}$ and we use a rate of
transmission of $r_{i,j} \log \rho$. This corresponds to a source
coding rate of $(b/N_t) r_{i,j} \log \rho$. The order in which the
source coded bits are mapped to the transmission layers is
important. The source coded bits are successively mapped on to the
transmitted layers from top left to bottom right going along each
row. That is, in the order $(1,1), \ldots, (1,N_t), (2,1), \ldots,
(2,N_t), \ldots, (N_s,N_t)$ (see Fig. \ref{fig:box}). The decoding
proceeds in the same order and when a layer cannot be decoded, the
source is reconstructed using all the layers that have been
successfully decoded up to that layer.

Let $\overline{r}_{(i-1)N_t+j} \log \rho = (b/N_t) \left(
\sum_{p=1}^{i-1} \sum_{q=1}^{N_t} r_{p,q}
+ \sum_{q=1}^{j-1} r_{i,q} \right)
\log \rho$
denote the cumulative source
coding rate up to the $(i,j)$th layer.
%Let $P_{i,j}$ denote the
%probability that the $(i,j)$th layer is in outage. Then,
%\begin{equation}
%P_{i,j} \doteq \rho^{-d^\star(r_{i,j},\gamma_{i-1,j},\gamma_{i,j})}
%\end{equation}
%where the exponent is given by lemma \ref{lemma:dstar}.
%
%Let $D_{i,j}$ be the distortion when the $(i,j)$th block cannot be
%decoded (outage).
As in the broadcast scheme case, we can approximate the overall
distortion up to an exponential order by
\begin{equation}
D \doteq \sum_i \sum_j \rho^{-{
d^\star(r_{i,j},\gamma_{i-1,j},\gamma_{i,j}) +
\overline{r}_{(i-1)N_t+j}}} + \rho^{-\overline{r}_{N_sN_t+1}}.
\end{equation}

Let $r$ and $\gamma$ denote the set of $r_{i,j}$'s and
$\gamma_{i,j}$'s. For a given $r,\gamma$, the overall exponent of
the scheme $a(b,r,\gamma)$ is then,
\begin{equation}
a(b,r,\gamma) =
\min_{i,j}(d^\star(r_{i,j},\gamma_{i-1,j},\gamma_{i,j}) +
\overline{r}_{(i-1)N_t+j},\overline{r}_{N_sN_t+1}) .
\end{equation}

The best achievable exponent with this scheme $a(b)$ is then given
by
\begin{equation}
a(b) = \max_{r,\gamma} a(b,r,\gamma).
\label{eqn:boxexp}
\end{equation}

If we allow for change in the bandwidth allocated to each layer,
then both BS and LSBLEND become special cases of this scheme and therefore
the exponent obtained from the maximization should be better than those
reported earlier. We will now show that for the
distortion SNR exponent, even with fixed bandwidth allocation to each layer,
the Box scheme can be designed to perform at least as well as BS and LSBLEND.

\begin{claim}
The Box scheme with $N_s$ superposition layer and $N_t$ time layers
has a distortion SNR exponent that is at least as good as that of the broadcast
scheme with $N_s$ layers.
\label{lemma:BSvsBox}
\end{claim}
\begin{proof}
  Let the optimal power allocation for the broadcast scheme by
$r_i$, $\gamma_i$. The exponent corresponding to the $i$th broadcast
layer is
$a_{BS}(i) = b \sum_{j=1}^{i-1}{r_j} + d^{*}(r_i,\gamma_{i-1},\gamma_i)$.
Now consider the Box scheme where the power allocation to the
$(i,j)$th layer $\gamma_{i,j}$ is set to $\gamma_i$ and the rate
$r_{i,j}$ = $r_i$.
Then
$a_{Box}(i,j) =  b \sum_{j=1}^{i-1}{r_j} + \frac{b}{N_t}(j-1) r_i +  d^{*}(r_i,\gamma_{i-1},\gamma_i)$.
Clearly $\frac{b}{N_t} \sum_{i,j} r_{i,j} = b \sum r_i$
and $a_{Box}(i,j) \geq a(i)$. Therefore, $\min_{i,j} (a_{Box}(i,j), \frac{b}{N_t}\sum_{i,j} r_{i,j})  \geq \min_i (a_{BS}(i),b \sum r_i) $.
In this case it is actually equal but if we optimize the power allocation
of the box scheme it could possibly improve on the exponent.
\end{proof}
\begin{claim}
In the limit as $N_t \rightarrow \infty$, the Box scheme has
 a distortion SNR exponent that is at least as good as that of LSBLEND.
\end{claim}
\begin{proof}
Consider the case when $(b-b_k)/N_t = b/N_{t,Box}$
where $N_t, N_{t,Box}$ are positive integers.
Consider a power and rate allocation for the box scheme that is
identical to the LSBLEND scheme for first $N_t$ time layers.
That is, the first $N_t$ time layers have no superposition layers
and the rate is identical to that of LSBLEND. For the remaining
$N_{t,Box} - N_t$ layers we allocate  power and rate with
the procedure used in lemma \ref{lemma:BSvsBox} and therefore its
contribution to the exponent is identical to the contribution of
the broadcast layer of LSBLEND.
Therefore, this has an exponent that is identical to
that of LSBLEND. Again, by optimizing the power and rate allocation
of the box scheme we could possibly improve the exponent.

For the case when $\frac{b}{b-b_k}$ is irrational, the result still
holds because the achievable exponent with LSBLEND and Box scheme
is a continuous function of $b$.
\end{proof}

The maximization in (\ref{eqn:boxexp}) is
difficult to perform analytically and very quickly becomes difficult
to perform even numerically. The procedure described in Algorithm
\ref{algo:greedy} has been used to find a suboptimal set of $r,
\gamma$. Remarkably, it turns out that for a range of $b$, this
achieves performance very close to the informed transmitter upper bound
$a_{IT}(b)$. Furthermore, for the considered examples, this scheme
performs nearly as well as currently known schemes for all $b$ while
it is strictly better for some range of $b$.
\begin{algorithm}
\begin{description}
\item[\textbf{Step 1:}]\ \ Initialization - Set $\gamma_{0,j}=1\ \forall j$ and $\overline{r}_{1}=0$.
\item[\textbf{Step 2:}]\ \ For $i = 1$ to $N_s$
\item[\textbf{Step 3:}]\ \ For $j = 1$ to $N_t$
\item[\textbf{Step 4:}]\ \ If $MN \gamma_{i-1,j} + \overline{r}_{(i-1)N_t+j} < d$,
set $\gamma_{i,j} = \gamma_{i-1,j}$ and goto step 10.
\item[\textbf{Step 5:}]\ \ Find smallest $k_{i,j} \in \{0, 1,\ldots, m-1\}$ such that
$0 \leq  \delta_{i,j} < \gamma_{i-1,j}$ where
  $\delta_{i,j} = ((M-k_{i,j})(N-k_{i,j})\gamma_{i-1,j} + \overline{r}_{(i-1)N_t+j}-d)/(M+N-1-2k_{i,j})$.
%\\ (It can be shown that such a $k$ always exists).
\item[\textbf{Step 6:}]\ \  If $i = N_s$ set $\gamma_{i,j}=0$ else set $\gamma_{i,j} = \gamma_{i-1,j}-\delta_{i,j}$
\item[\textbf{Step 7:}]\ \  Set $r_{i,j} =  k_{i,j}(\gamma_{i-1,j}-\gamma_{i,j})+ \delta_{i,j}$.
\item[\textbf{Step 8:}]\ \  Update $\overline{r}_{(i-1)N_t+j+1} =\overline{r}_{(i-1)N_t+j}+ (b/N_t)r_{i,j}$
\item[\textbf{Step 9:}]\ \  If $\overline{r}_{(i-1)N_t+j+1} > d$, exponent $d$ is achievable. return.
\item[\textbf{Step 10:}]\ \ \ End of $j$ loop
\item[\textbf{Step 11:}]\ \ \  End of $i$ loop
\item[\textbf{Step 12:}]\ \ \  Exponent $d$ is not achievable using this scheme. return.
\end{description}
\caption{Algorithm to check if an exponent $d$ is achievable using
the proposed scheme} \label{algo:greedy}
\end{algorithm}

For each $(i,j)$ if we fix $k_{i,j} \in \{0, 1,\ldots, m-1\}$ and let $r_{i,j} =
k_{i,j}(\gamma_{i-1,j}-\gamma_{i,j})+ \delta_{i,j}$ where $0 \leq
\delta_{i,j} < (\gamma_{i-1,j}-\gamma_{i,j})$, then, as before, the
problem of finding the optimal exponent reduces to a linear program
and hence by solving it for different $k_{i,j}$'s we would expect to
find an exponent that is better than that obtained using Algorithm
\ref{algo:greedy}. However, in Step 4 of the algorithm, notice that
we skip a layer if it is not possible to allocate a non zero rate.
Therefore, this layer is never in outage. However, in the linear
program, if we use lemma \ref{lemma:dstar} to compute
$d^{*}(0,\gamma_{i-1,j},\gamma_{i-1,j})$ we get 0 which means this
layer is always in outage. Therefore, to obtain the optimal exponent, we
will need to allow for a layer to be skipped in addition to allowing
for different values of $k$ for that layer. The complexity thus
grows as $(M+1)^{N_sN_t}$.

The achievable exponent with this scheme increases monotonically
with $N_S$. Interestingly, the achievable exponent with this scheme
may not increase monotonically with $N_t$.

We also considered the following variations, which provide some
gain for finite number of layers. However, the gain diminishes as
the number of layers increases.

\subsubsection{Adding an Analog Layer}\label{sec:HDA}

In this scheme, we start allocating rate and power levels to the
layers as in Algorithm \ref{algo:greedy}. Let us denote by ${\cal
A}_{i,j} = \{(p,q):p\leq i-1 \hbox{ AND } q \leq N_t\} \cup \{(i,q): q
< j \}$ the set of all layers for which a rate and power allocation
has been found during the $(i,j)$th stage of the algorithm. Let
$\Xm^a_{i,j}$ denote the analog quantization error in quantizing the source
using $\overline{r}_{(i-1)N_t+j} \log \rho$ bits. We check if
at least $\lceil \frac{N_t}{b m} \rceil$ layers are still available
in ${\cal A}_{i,j}^c$ to transmit the analog quantization error
such that the desired exponent can be achieved. If this is
possible, we stop there and this becomes the overall transmission
scheme. Otherwise, we allocate a power level $\gamma_{i,j}$  and
rate $r_{i,j}$ corresponding to the $(i,j)$th layer as before and
continue to the next layer. Note that this contains the HLS schemes
of \cite{caire05,erkip-IT} as a special case (when $N_s=1$).

\subsubsection{Ordering the layers based on available power}
 In this variation, we allocate rate and power as in Algorithm
\ref{algo:greedy} but the order of selecting the layers is
not sequential. At any stage
of the algorithm, we select the time layer with the maximum available
power (total power minus power already allocated to superposition layers
in that time layer). A new superposition layer is added to this layer
with power and rate allocation as specified in Algorithm \ref{algo:greedy}.
%This order is different from the sequential order since the power
%allocated to the $i$th superposition layers is different for the
%different time layers.
Note that this is the order in which the successive refinement information
from the source coder is filled and therefore the decoder should
decode the layers in this order.

\section{Extensions to Multiple Block Fading Channels}
\label{sec:generalL} In this section we extend the results derived
for the MIMO channel to the $L$-block fading MIMO channel.
We assume that $K$ source samples are transmitted over $L$ blocks
of length $T/L$ each. The fading coefficient for the different blocks
are independent.

\begin{lemma}
\label{lemma:dstarL} If the multiplexing gain in the $i$th layer of
the broadcast scheme is $r_i = \frac{kL+a}{L} (\gamma_{i-1} -
\gamma_{i}) + \delta $ where $k \in \{0,1,\ldots,m-1\}$, $a \in
\{0,1,\ldots,L-1\}$ and $0 \leq \delta < \frac{\gamma_{i-1} -
\gamma_{i}}{L}$, then, the achievable diversity in the $i$th layer
of the broadcast scheme, assuming that the message transmitted in
the previous layers is available at the receiver, is given by
\begin{eqnarray}
\nonumber
d^\star(r_{i},\gamma_{i-1},\gamma_{i}) =L(m-k) (n-k) \gamma_{i-1} - (m+n-1-2k)(a\gamma_{i-1} + L\delta).
\label{eqn:dstarL}
\end{eqnarray}
That is, if
$X = \frac{1}{\sqrt \rho}(\sum_{i=1}^{N_s} \sqrt{\rho^{\gamma_{i-1}} -
\rho^{\gamma_i}}X_i + \sqrt{\rho^{\gamma_{N_s}}}N_1)$,
where $X_i,N_1 \sim {\cal N}(0,I^{M\times M})$,
$Y_l = \sqrt{\frac{\rho}{M}}{\bf H_l} X + N$ for $l = 1, \ldots, L$,
and ${\cal A}_i = \{H_1,\ldots,H_L:\frac{1}{L}\sum_{l=1}^{L} I(X_i;Y_l|{\bf H_1}=H_1,\ldots,{\bf H_L}=H_L,X_1,\ldots X_{i-1}) < r_i \log \rho)$ denotes the outage event set then $P({\cal A}_i) \doteq
\rho^{-d^\star(r_{i},\gamma_{i-1},\gamma_{i})}$.
\end{lemma}
\begin{proof}
\begin{equation}
P({\cal A}_i) = P\left(\frac{1}{L}\sum_{l=1}^{L}\log \frac{\det (I +
\frac{1}{M}\rho^{\gamma_{i-1}} {\bf H_l} {\bf H_l}^H)} {\det (I +
\frac{1}{M}\rho^{\gamma_i} {\bf H_l} {\bf H_l}^H)} < r_i \log \rho\right).
\end{equation}

Let $\lambda_{1,l},\ldots,\lambda_{m,l}$ denote the non-zero ordered
eigenvalues of ${\bf H_l} {\bf H_l}^H)$ with $\lambda_{1,l} \leq
\lambda_{2,l} \leq \ldots \lambda_{m,l}$. As in \cite{zheng-tse},
let $\alpha_{j,l} = - \frac{\log \lambda_{j,l}}{\log \rho}$.
Therefore, $\alpha_{1,l} \geq \alpha_{2,l} \geq \ldots
\alpha_{m,l}$. At high SNR, $P({\cal A}_i) \doteq P({\cal A'})$ where
\begin{equation}
{\cal A'}= \left\{\alpha: \frac{1}{L}\sum_{l=1}^{L}\sum_{j=1}^{m}\left(
(\gamma_{i-1} - \alpha_{j,l})^+ - (\gamma_{i}-\alpha_{j,l})^+
\right) < r_i \right\}.
\end{equation}

Following in the footsteps of \cite{zheng-tse}, we observe that the
outage probability is given by $P({\cal A}_i) \doteq
\rho^{-d^\star(r_i,\gamma_{i-1},\gamma_{i})}$ where
\begin{equation}\label{eqn:dstar1L}
d^\star(r_i,\gamma_{i-1},\gamma_{i}) = \inf_{{\cal A'} \cap
\alphav^+} \sum_{j=1}^{m} \sum_{l=1}^{L}(2j-1+n-m) \alpha_{j,l}.
\end{equation}

For $r_i = \frac{kL+a}{L}(\gamma_{i-1}-\gamma_i) + \delta$ where $k
\in [0,1,\ldots,m-1]$, $a \in [0,1,\ldots,L-1]$, and $0 \leq \delta
< \frac{\gamma_{i-1}-\gamma_i}{L}$, the infimum in
(\ref{eqn:dstar1L}) occurs when $\alpha = \alpha^{*}$ where
\begin{equation}\alpha_{j,l}^{*}=\left\{
       \begin{array}{ll}
           \gamma_{i-1}, & 1 \leq j < m-k;\\
            \gamma_{i-1} , & j = m-k, 1 \leq l < L-a;\\
           \gamma_{i-1} - L\delta, & j = m-k, l = L-a;\\
            0, &  j=m-k, L-a < l \leq L ; \\
            0, &  m-k < j \leq m
       \end{array}\right.
\label{eqn:alphaInf}
\end{equation}
%%Note that since $\delta < \gamma_{i-1}-\gamma_i$ and $0 \leq
%%\gamma_i \leq \gamma_{i-1}$, we have $0 \leq \delta < \gamma_{i-1}$
%%and therefore $\alpha_{m-k} > 0$.
%%On substituting in
%%(\ref{eqn:dstar1L}) and simplifying we get the desired result.
Hence,
\begin{eqnarray}
\nonumber \lefteqn{d^\star(r_i,\gamma_{i-1},\gamma_{i})}
\\ \nonumber
 &=& \sum_{j=1}^{m-k-1}
(2j-1+n-m) L\gamma_{i-1}
 + (2(m-k)-1+n-m) ((L-a)\gamma_{i-1} -
\delta)
\\ \nonumber
& = & \sum_{j=1}^{m-k} (2j-1+n-m) L\gamma_{i-1}
- (2(m-k)-1+n-m) (a\gamma_{i-1}+L\delta)
\\ \nonumber
& = & L(m-k)(2\frac{m-k+1}{2}-1+n-m) \gamma_{i-1}
- (m+n-1-2k)(a \gamma_{i-1}+L\delta).
\end{eqnarray}
This then gives the desired result.
\end{proof}

\begin{thm}
\label{thm:broadcastL} The broadcast scheme achieves a distortion
SNR exponent of $\frac{kL+a+1}{L}b$, $k \in \{0,1,\ldots,m-1\}$, $a \in
\{0,1,\ldots,L-1\}$ for $ \frac{L(M-k)(N-k)-(a+1)(M+N-1-2k)}{kL+a+1}
< \frac{b}{L} < \frac{L(M-k)(N-k)-a(M+N-1-2k)}{kL+a+1} $ with
power and rate allocation
\begin{equation}
\gamma_i  = \alpha^i
\end{equation}
and
\begin{equation}
r_i =  \frac{kL+a+1}{L} (\gamma_{i-1} - \gamma_{i}) - \epsilon
\end{equation}
where
\begin{equation} \alpha = 1+a+\frac{b\frac{kL+a+1}{L}
-L(M-k)(N-k)}{M+N-1-2k}
\end{equation}
for arbitrarily small $\epsilon > 0$.
\end{thm}
\begin{proof}
  The power and rate allocation policy can be derived in a manner similar
to that in Theorem \ref{thm:broadcast}. Here we will just verify that the
specified rate and power allocation policy indeed gives the specified
exponent.

We first note that for $ \frac{L(M-k)(N-k)-(a+1)(M+N-1-2k)}{kL+a+1}
< \frac{b}{L} < \frac{L(M-k)(N-k)-a(M+N-1-2k)}{kL+a+1}$, $0 < \alpha <1$
and therefore the specified rate and power allocation is a valid assignment.

As in Theorem \ref{thm:broadcast} the distortion SNR exponent is given by
\begin{equation}
  a_{BS} = \min (b \sum_j r_j, a(1),\ldots, a(i),\ldots)
\end{equation}
where
\begin{equation}
a(i) = b \sum_{j=1}^{i-1} r_j + d^{*}(r_i, \gamma_{i-1}, \gamma_i).
\label{eqn:bslayern}
\end{equation}
We have
\begin{equation}
\lim_{i \rightarrow \infty} b \sum_{j=1}^{i} {r_j} = b\frac{kL+a+1}{L} (1 -\lim_{i \rightarrow \infty} \gamma_{i})= b\frac{kL+a+1}{L}.
\end{equation}
Furthermore, from (\ref{eqn:bslayern}) and lemma \ref{lemma:dstarL} we have
\begin{eqnarray}
\nonumber
  a(i) &=& b \frac{kL+a+1}{L}(1-\gamma_{i-1}) + L(m-k)(n-k)\gamma_{i-1} - (m+n-1-2k)((a+1)\gamma_{i-1}-\gamma_i) \\
\nonumber
&=& b \frac{kL+a+1}{L} + (m+n-1-2k) \gamma_{i} \\\nonumber&& \qquad + \gamma_{i-1}\left(L(m-k)(n-k)-(m+n-1-2k)(a+1) - b\frac{kL+a+1}{L}\right)
\\\nonumber
&=& b \frac{kL+a+1}{L} + (m+n-1-2k) (\gamma_{i} -\alpha \gamma_{i-1}) =b \frac{kL+a+1}{L}.
\end{eqnarray}
Therefore, $a_{BS} = b \frac{kL+a+1}{L}$.
\end{proof}

By comparing with the upper bound, we observe that the broadcast
scheme achieves the optimal exponent of $m b$ for $b <
\frac{n-m+1}{m}$ and $MNL$ for $b > MNL^2$. This has been shown
earlier for the $M=N=1$ case in \cite{erkip-ISIT} and for the $L=1$
case in \cite{erkip-IT}.

%\begin{thm}
%Let $c_{kL+a} = L(m+n-1-2k) \log\left( \frac{(k+a/L)}{(k
%+(a-1)/L)}\right)$. Consider the time layering scheme with a
%broadcast layer of bandwidth $b_k = \frac{(m-k)(n-k)L -
%(a-1)(n+m-1-2k)}{k+\frac{a}{L}}$ at the end. Let
%$\sum_{i=k_1L+a_1}^{kL+a-1} c_i < b-b_k <
%\sum_{i=k_1L+a_1-1}^{kL+a-1} c_i$ where $k_1 \in \{0,1,\ldots,m-1\}$
%and $a_1 \in \{0,1,\ldots,L-1\}$. The distortion exponent is then
%given by
%\begin{equation}
%a_{kL+a}(b) = (m-k_1)(n-k_1)L - (r_1 - k_1)L(n+m-1-2k_1)
%\end{equation}
%where \begin{equation} r_1 = (k_1+\frac{a_1}{L}) e^{
%-\frac{b-b_{kL+a} - \sum_{i=k_1L+a_1}^{kL+a-1}c_i}{L(m+n-1-2k_1)}}.
%\end{equation}
%\end{thm}

\begin{thm}
Let $c_{kL+a} = L(m+n-1-2k) \log\left( \frac{kL+a+1}{kL+a}\right)$
for $k \in \{0,1,\ldots,m-1\}$ and $a \in \{0,1,\ldots,L-1\}$.
Consider the time layering scheme with a broadcast layer of
bandwidth $b_{kL+a} = \frac{(m-k)(n-k)L^2 - aL(n+m-1-2k)}{kL+a+1}$
at the end. Let $k_1$ and $a_1$ be such that
$\sum_{j=k_1L+a_1+1}^{kL+a-1} c_j < b-b_{kL+a} <
\sum_{j=k_1L+a_1}^{kL+a-1} c_j$ where $k_1 \in \{0,1,\ldots,m-1\}$
and $a_1 \in \{0,1,\ldots,L-1\}$. The distortion SNR exponent is then
given by
\begin{equation}
a_{kL+a}(b) = (m-k_1)(n-k_1)L - (r_1 - k_1)L(n+m-1-2k_1)
\end{equation}
where \begin{equation} r_1 = (k_1+\frac{a_1+1}{L}) e^{
-\frac{b-b_{kL+a} -
\sum_{i=k_1L+a_1+1}^{kL+a-1}c_i}{L(m+n-1-2k_1)}}.
\end{equation}
\end{thm}
\begin{proof}
  The proof is similar to that of Theorem \ref{thm:lsblend}
and is skipped here.
\end{proof}

\section{Numerical Results}
\label{sec:results} 
\begin{figure}
\begin{center}
\includegraphics*[width = 5in]{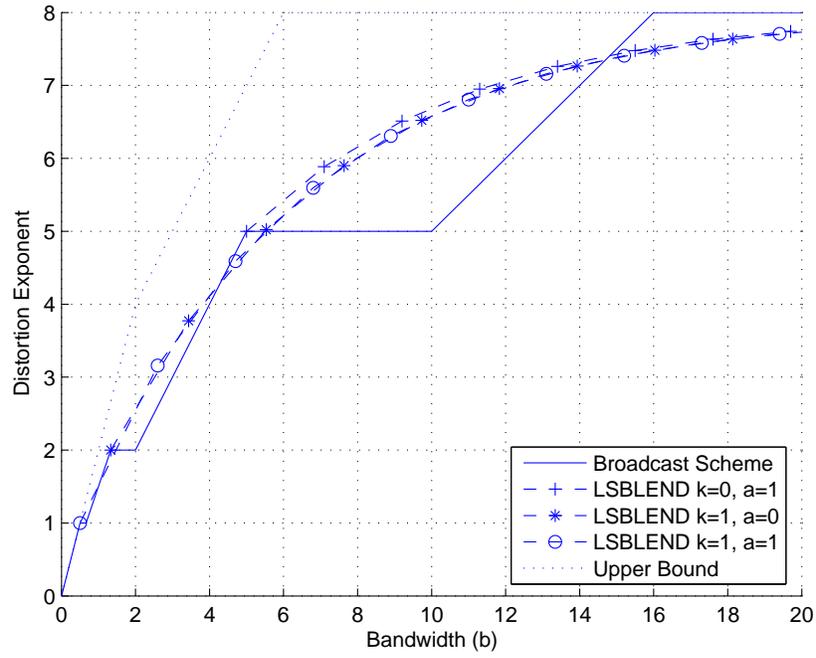}
\caption{Achievable distortion
SNR exponent for $M=N=2$} \label{results2x2}
\end{center}
\end{figure}

\begin{figure}
\begin{center}
\includegraphics*[width = 5in]{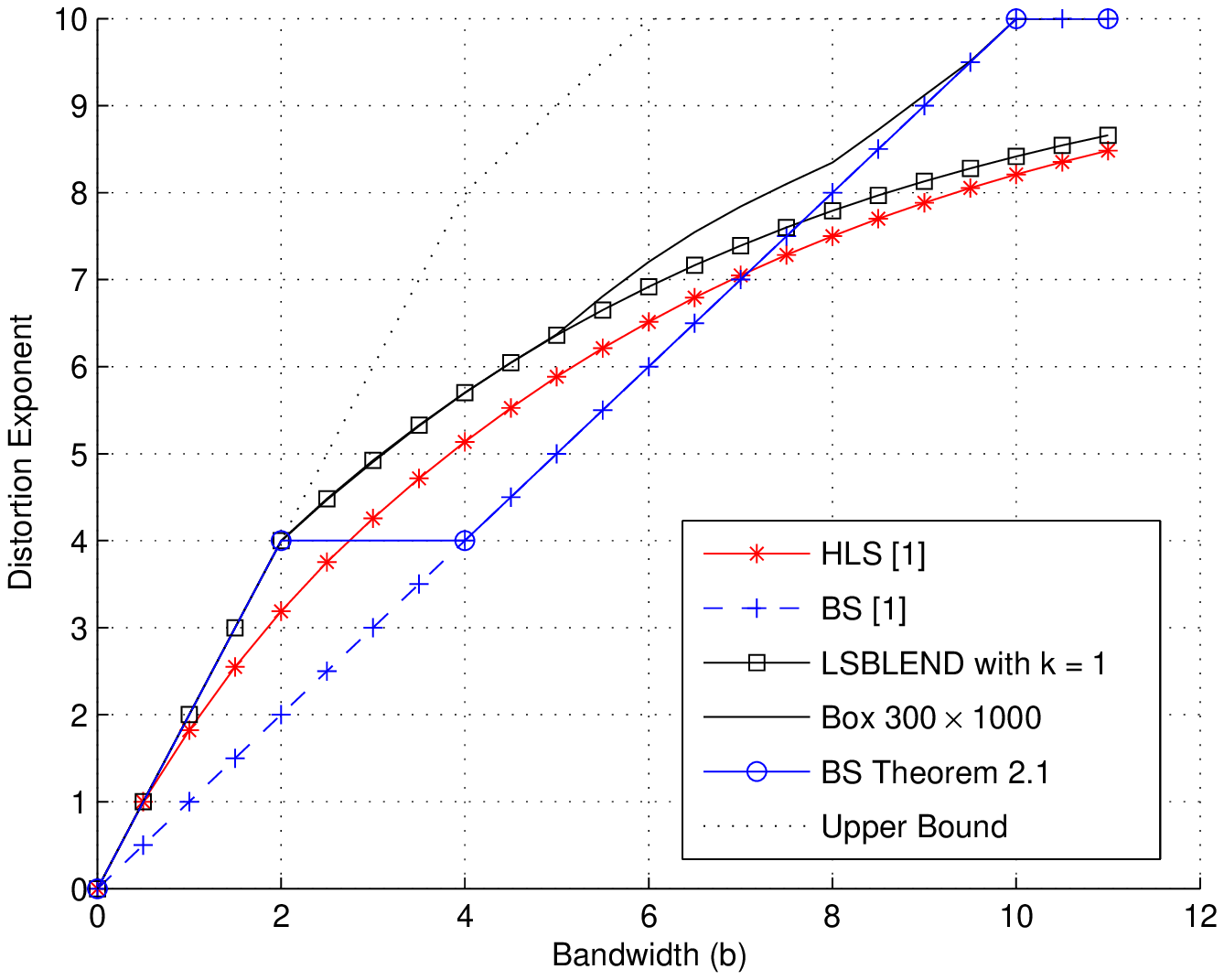}
\caption{Achievable distortion SNR exponent for $M=2, N=5$}
\label{results2x5}
\end{center}
\end{figure}

The achievable exponent using the proposed
hybrid layering scheme along with that achievable by the HLS and
broadcast schemes of \cite{erkip-IT} are shown in
Fig.~\ref{results2x2} and Fig. \ref{results2x5}. Note that
the proposed schemes outperforms the schemes in \cite{erkip-IT} for
all $b$, making this the best known achievable distortion SNR exponent.

The optimal exponent can be obtained for all $b < (n-m+1)/m$ using
the purely digital scheme in Section~\ref{sec:bs}. This is the first
time a scheme has been shown to obtain the optimal exponent for $1/m
< b < (n-m+1)/m$. Since the scheme in Section~\ref{sec:bs} is a
special case of the scheme in Section~\ref{sec:box}, the optimal
exponent is achieved in this region by the scheme in
Section~\ref{sec:box} as well.

\begin{figure}
\begin{center}
\includegraphics*[width = 5in]{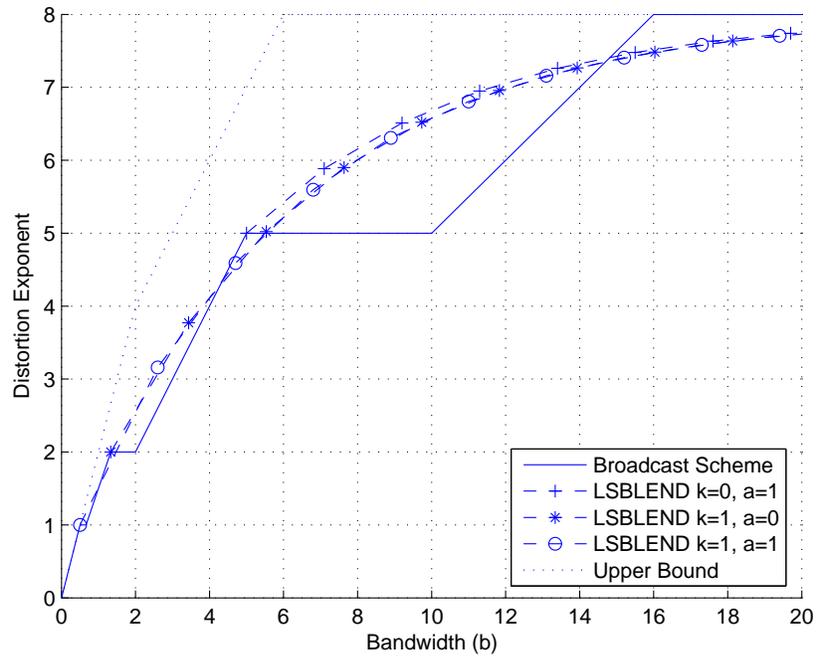}
\caption{Achievable distortion SNR exponent for $M=N=L=2$}
\label{results2x2x2}
\end{center}
\end{figure}

A plot of the distortion SNR exponent for $M = N = L = 2$ is shown in
Fig.~\ref{results2x2x2}.

\begin{figure}
\begin{center}
\includegraphics*[width=5in]{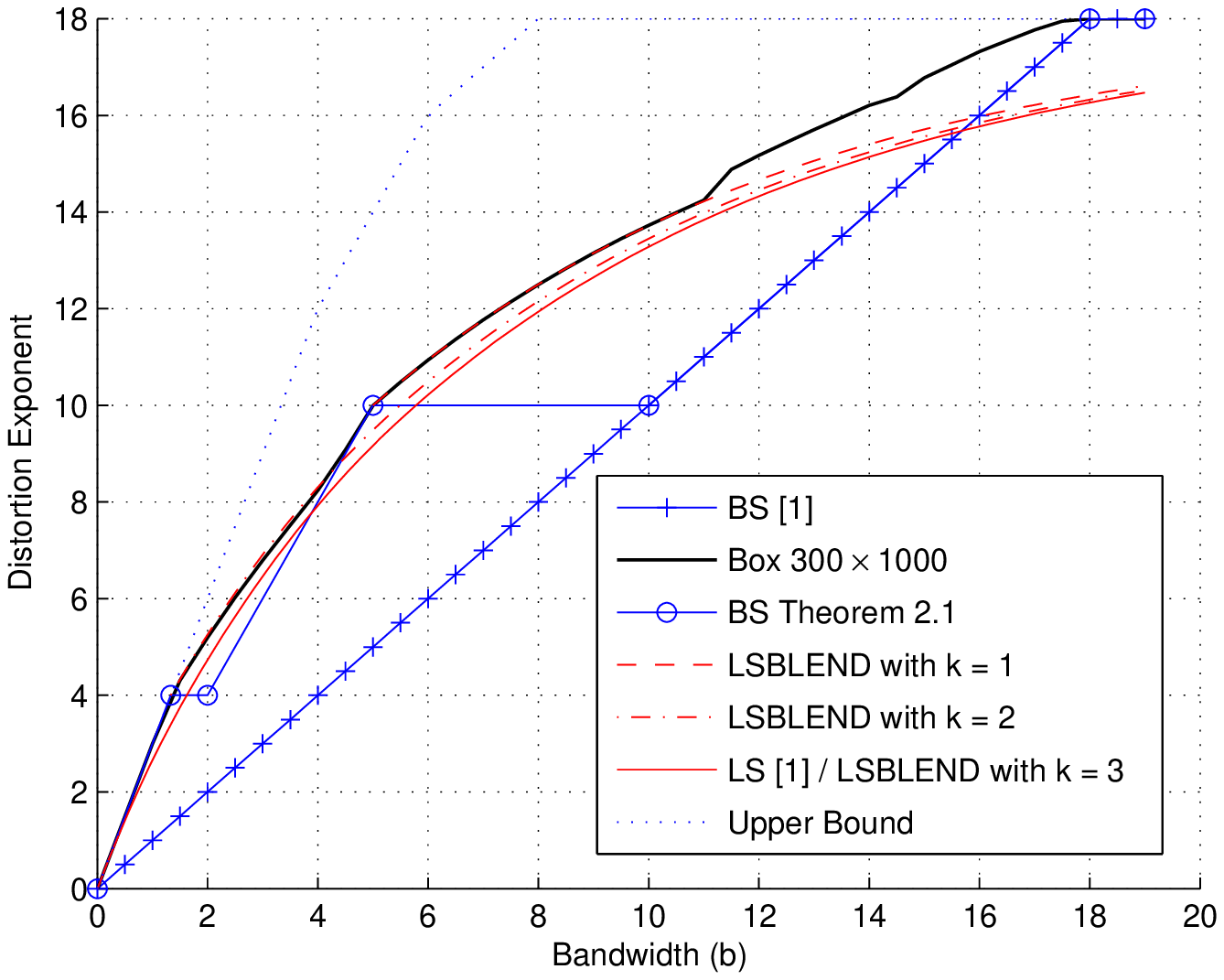}
\caption{Achievable distortion SNR exponent for $M=3, N=6$}
\label{results3x6b}
\end{center}
\end{figure}

%\begin{figure}
%\begin{center}
%\includegraphics*[width=5in]{schemes}
%%\epsfxsize=4in \epsffile{2by3.eps}
%\caption{Different Schemes} \label{schemes}
%\end{center}
%\end{figure}

\section{Conclusion}
\label{sec:conclusion} We have proposed layering schemes for
transmitting a discrete time analog source over a block fading
MIMO channel.
Achievable distortion SNR exponent using carefully selected rate and
power allocation policies for these scheme have been studied. The
achievable distortion SNR exponent obtained using these schemes is better
than those reported in \cite{caire05,erkip-IT} making
this the best known
distortion SNR exponent so far. Particularly, the optimal exponent is
obtained for $b < (n-m+1)/m$ and $b > mnL^2$. We believe this is a
new and surprising result. Our current research focusses on
optimizing the rate and power allocation of these schemes.

%\section{notes}
%\begin{equation}
%(m-k)(n-k) - (m+n-1-2k)  = (m-k-1)(n-k-1)
%\end{equation}

\appendix
\subsection{Optimality of equating exponents for the BS}
\label{sec:appBS}
The exponent $a(b)$ is given by the following optimization problem
\begin{eqnarray*}
&a(b)& = - \min_{a,\gamma_1,\ldots,\gamma_{N_s}} -a \\
&&\mbox{subject to:}\\
&&C_i = a - b(k+1)(1-\gamma_{i-1}) - (m-k)(n-k)\gamma_{i-1}+(m+n-1-2k)(\gamma_{i-1} - \gamma_i) \leq 0 \\ &&\qquad \qquad {\mbox{for }} i = 1,2,\ldots,N_s;\\
&& C_{N_s+1}=a - b(k+1)(1-\gamma_{N_s}) \leq 0; \\
&& \gamma_{i+1} \leq \gamma_{i}; \ \ \gamma_{N_s} \geq 0; \ \ \gamma_{0} = 1.
\end{eqnarray*}
We solve this optimization problem by ignoring the constraints
$\gamma_{i+1} \leq \gamma_{i}$ and $\gamma_{N_s} \geq 0$. Any solution
then is an upper bound on $a(b)$. Furthermore, if the solution satisfies
the ignored constraints then the solution yields $a(b)$.
Consider the function $F = -a + \sum_{i=1}^{N_s+1} \lambda_i C_i  $.
Setting $dF/d\gamma_i = 0$ we have for $i = 1$ to $N_s -1$
\begin{eqnarray*}
\frac{dF}{d\gamma_i} =  -\lambda_i (m+n-1-2k) + \lambda_{i+1} (b(k+1) - (m-k)(n-k) + (m+n-1-2k)) = 0 \\
\Rightarrow \lambda_i = \frac{b(k+1)- (m-k)(n-k) + (m+n-1-2k)}{m+n-1-2k} \lambda_{i+1} = \alpha \lambda_{i+1}.
\end{eqnarray*}
For $i = N_s$
\begin{equation}
\frac{dF}{d\gamma_{N_s}} =  -\lambda_{N_s} (m+n-1-2k) + \lambda_{N_s+1} b(k+1) = 0.
\end{equation}
Therefore, we have
\begin{equation}
\lambda_i = \alpha^{N_s-i} \frac{b(k+1)}{m+n-1-2k} \lambda_{N_s+1}.
\end{equation}
Setting $dF/da = 0$ we have
\begin{equation}
-1 + \sum \lambda_i = 0.
\end{equation}
We are interested in the region $b > (m-k-1)(n-k-1)/(k+1)$ and
therefore $\alpha > 0$. So all $\lambda$'s are strictly positive.
Therefore, from the KKT conditions, it follows that the optimal
solution satisfies $C_i = 0$ for $i = 1$ to $N_S+1$.

\subsection{Optimality of equating exponents for LSBLEND}
\label{sec:appLSBLEND}
The exponent $a(b)$ is given by the following optimization problem
\begin{eqnarray*}
&a(b)& = - \min_{a,r_1,\ldots,r_{N_t}} -a \\
&&\mbox{subject to:}\\
&&C_i = a - \frac{b-b_k}{N_t} \sum_{j=0}^{i-1}r_j - d^{*}(r_i)\leq 0 \qquad {\mbox{for }} i = 1,2,\ldots,N_t;\\
&& C_{N_t+1}=a -\frac{b-b_k}{N_t}\sum_{j=0}^{N_t}r_j + b_k(k+1) \leq 0; \\
&& r_i \geq 0;\ \ r_i < m; \ \ r_0 = 0.
\end{eqnarray*}
As in Appendix \ref{sec:appBS} we ignore the constraints $0 < r_i < m$
and consider the function $F = -a + \sum_{i=1}^{N_t+1} \lambda_i C_i  $.
Setting $dF/dr_i = 0$ we have for $i = 1$ to $N_t$
\begin{equation}
\frac{dF}{dr_i} = -\frac{b-b_k}{N_t} \sum_{j = i+1}^{N_t+1}\lambda_j  - \lambda_i \frac{d}{dr_i}(d^{*}(r_i)) = 0.
\end{equation}
Note that $\frac{b-b_k}{N_t} > 0$ and $\frac{d}{dr_i}(d^{*}(r_i)) < 0$.
Starting from $i = N_t$ and solving recursively for $\lambda_i$ in terms
of $\lambda_{N_t+1}$ we observe that the $\lambda_i$'s are of form
$\alpha_i \lambda_{N_s+1}$ where $\alpha_i > 0$. By setting $dF/da = 0$ we have
$\sum_{i=1}^{N_t+1} \lambda_i = 1$.
Therefore, $\lambda_i >  0$ for all $i$ and hence, from the KKT conditions,
we conclude that the optimal solution satisfies $C_i = 0$ for
$i = 1$ to $N_t+1$.

%Let $\beta_i = \sum_{i}^{N_t+1} \lambda_i$. We have
%\begin{eqnarray}
%\frac{dF}{dr_i} = -\frac{b-b_k}{N_t} \beta_{i+1}  - (\beta_i-\beta_{i+1}) \frac{d}{dr_i}(d^{*}(r_i)) = 0\\
%\Rightarrow \beta_i = \frac{\frac{b-b_k}{N_t} - \frac{d}{dr_i}(d^{*}(r_i))}{\frac{d}{dr_i}(d^{*}(r_i))} \beta_{i+1}
%\end{eqnarray}
%Note that $\frac{b-b_k}{N_t} > 0$ and $\frac{d}{dr_i}(d^{*}(r_i)) < 0$.
%Therefore $\beta_i$ is of form $\beta_{i+1} \alpha_i $ where $\alpha_i > 1$.
%By setting $dF/da = 0$ we have $\beta_1 = \sum_{i=1}^{N_t+1} \lambda_i = 1$. Therefore
%$\beta$'s form a decreasing positive sequence and hence $\lambda_i > 0$ for all $i$.
%Therefore for the optimal solution
%$C_i = 0$ for $i = 1$ to $N_t+1$. In this case too we will later see that the solution obtained for the relaxed problem satisfies the other constraints.

\bibliography{journal}
\bibliographystyle{unsrt}

\end{document}